\documentclass[aps,prd,onecolumn,superscriptaddress,preprintnumbers,nofootinbib,a4paper,10pt]{revtex4-2}

\usepackage[utf8]{inputenc} 
\usepackage[T1]{fontenc}

\usepackage[portuguese,english]{babel}

\usepackage{amsfonts,amsmath,bbm,latexsym,amssymb,amsthm}

\newcommand{\Z}[1]{\ensuremath{\mathbb{Z}_{#1}}} 
\newcommand{\Zbold}[1]{\mathbf{Z_{#1}}}  

\usepackage{physics}   

\usepackage{slashed}

\usepackage{graphicx}

\usepackage{tensor}
\usepackage{booktabs,multirow}
\usepackage{cancel}
\usepackage{parskip}
\usepackage{float}
\usepackage{xcolor}
\usepackage{tabularx}
\usepackage{comment}
\usepackage[all]{xy}   

\usepackage{epsfig}
\usepackage{epstopdf}
\usepackage[normalem]{ulem}
\usepackage{soul}
\usepackage{amsfonts}

\usepackage{tikz}
\usepackage{tikz-feynman}

\usepackage[font=small,justification=raggedright,singlelinecheck=false]{caption}

\usepackage[bookmarksdepth=5]{hyperref} %
\definecolor{customblue}{RGB}{23, 125, 214}
\hypersetup{
    colorlinks=true,
    citecolor=customblue,
    linkcolor=black,
    filecolor=magenta,   
    urlcolor=cyan,
    pdfborder={0 0 0},
}
\usepackage{cleveref}

\begin{document}

\begin{flushright}
KA-TP-06-2026 \\
IPPP/26/19\\
\end{flushright}
\title{Machine Learning insights on the $\Z3$ 3HDM with Dark Matter}

\author{Fernando Abreu de Souza}
\email{abreurocha@lip.pt}
\affiliation{LIP -- Laborat\'orio de Instrumenta\c{c}\~ao e F\'isica Experimental de Part\'iculas, Escola de Ciências, Campus de Gualtar, Universidade do Minho, 4701-057 Braga, Portugal}
\author{Rafael Boto}%
\email{rafael.boto@tecnico.ulisboa.pt}
\affiliation{Institute for Theoretical Physics,
Karlsruhe Institute of Technology, \\
76128 Karlsruhe, Germany}
\author{Miguel Crispim Romão}
\email{miguel.romao@durham.ac.uk}
\affiliation{Institute for Particle Physics Phenomenology, Durham University, Durham DH1 3LE, UK}
\affiliation{LIP -- Laborat\'orio de Instrumenta\c{c}\~ao e F\'isica Experimental de Part\'iculas, Escola de Ciências, Campus de Gualtar, Universidade do Minho, 4701-057 Braga, Portugal}
\author{Pedro N. de Figueiredo}%
\email{pedro.m.figueiredo@tecnico.ulisboa.pt}
\affiliation{Departamento de
	F\'isica and CFTP,  Instituto Superior T\'ecnico,\\  \small\em
	Universidade de Lisboa, Av
	Rovisco Pais, 1, P-1049-001 Lisboa, Portugal}
\author{Jorge C. Romão}%
\email{jorge.romao@tecnico.ulisboa.pt}
\affiliation{Departamento de
	F\'isica and CFTP,  Instituto Superior T\'ecnico,\\  \small\em
	Universidade de Lisboa, Av
	Rovisco Pais, 1, P-1049-001 Lisboa, Portugal}

\begin{abstract}
We study a 3-Higgs Doublet Model (3HDM) with an imposed $\Z3$ symmetry, allowing for two Inert scalar doublets and one active Higgs doublet. The WIMP dark matter candidates correspond to two mass-degenerate states, $H_1$ and $A_1$, which possess opposite CP quantum numbers and can reproduce the correct relic density simultaneously with all theoretical and experimental constraints. 
We use state-of-the-art machine learning algorithms to probe the parameter space of the model by employing an Evolutionary Strategy augmented with Novelty Reward. We consider the multiple solutions to the condition that guarantees the vertex $ZH_iA_i \sim \cos(2\theta)$ to vanish, closing a channel that would otherwise be at odds with current direct detection bounds. For scenarios with $\theta=\tfrac{\pi}{4}$, we find viable dark matter candidates within two separate mass regimes, ranging from $ 50~\textrm{GeV} \leq m_{\textrm{DM}} \lesssim m_{\textrm{W}} $ and $386 \lesssim m_{\textrm{DM}} \lesssim 1000  \textrm{ GeV}$. The other possible solution considered only finds viable solutions in the low DM mass region. Moreover, we find it is possible to fulfill all the existing constraints while still obtaining values for the dark matter-higgs coupling of order $\sim \mathcal{O}(0.1)$. 

Exploring the model proved to be a challenging task, as there are regions which seem easy to explore when projected on a 2D plane, yet may be completely disconnected on the hypersurface supporting the valid points, given its non-convex and multi-dimensional nature. We consider new methods of prototype selection to seed new exploration runs, which allow for  efficient global scans over the parameter space.

\end{abstract}

\maketitle

\section{Introduction}

The Standard Model is our current best description of particle physics. In 2012, yet another of its postulates was proven with the discovery of the  $125\,\textrm{GeV}$ Higgs boson, that arises from a scalar doublet field, by the ATLAS \cite{ATLAS:2012yve}  and CMS \cite{CMS:2012qbp} collaborations.  Despite the success of the SM, there are still open problems that need to be addressed, such as Dark Matter (DM), which comprises about $85\%$ of all matter in the Universe~\cite{Planck:2013oqw}. Similar to the fermions, the number of bosons must be determined experimentally. 
Therefore, one may consider a N-Higgs Doublet Models (NHDM) approach to provide suitable DM candidates,
by extending the SM with additional scalar doublet fields with a preserved symmetry that stabilizes the DM candidates. 
Given its elusive nature, DM has yet to be detected directly and experiments place increasingly stringent bounds. The allowed parameter space of minimal extension models, such as the Inert Doublet Model (IDM)~\cite{Deshpande:1977rw,Ma:2006km,Barbieri:2006dq,LopezHonorez:2006gr} with a single additional doublet and a $\Z2$ symmetry, is thus becoming progressively more limited~\cite{Nezri:2009jd,Arina:2009um,LopezHonorez:2010eeh,LopezHonorez:2010tb,Arhrib:2013ela,Klasen:2013btp,Ilnicka:2015jba,Belyaev:2016lok,Eiteneuer:2017hoh,Kalinowski:2018ylg,Belyaev:2018ext,Ilnicka:2018def,Banerjee:2019luv,Banerjee:2021oxc,Ghosh:2021noq}. This motivates the study of larger models, since they provide richer phenomenology and can predict a DM candidate in regions already excluded in simpler models. 

In this work, we consider the possibility of a  three-Higgs-doublet model (3HDM) with an additional $\Z3$ symmetry~\cite{Ferreira:2008zy,Bento:2017eti,Aranda:2019vda,Das:2019yad,Boto:2021qgu,Boto:2023nyi,Romao:2024gjx,Chakraborti:2021bpy,Das:2022gbm,Coleppa:2025qst,Coleppa:2026ogl,Coleppa:2026fdj,Coleppa:2026esg}. Introduced with the notation I(2+1)HDM in Ref.~\cite{Aranda:2019vda}, the vacuum configuration is chosen that only the scalar doublet even under $\Z3$ acquires a vacuum expectation value (vev). The two other doublets, non-invariant under $\Z3$,  do not acquire a vev, achieving the vacua alignment $(0, 0, v)$ and an unbroken symmetry. Choosing the fermions to be invariant under $\Z3$, the two inert doublets are part of a stable dark sector. The symmetry being specifically $\Z3$ leads to two mass-degenerate DM states with opposite CP, distinct from a single $
\Z2$~\cite{Cordero-Cid:2016krd,Cordero:2017owj}, with only the lightest $\Z2$-odd scalar as the DM candidate, or a $\Z2\times\Z2$ construction \cite{Hernandez-Sanchez:2020aop,Hernandez-Sanchez:2022dnn,Boto:2024tzp,Romao:2025qag}, with two distinct dark sectors, each with one DM scalar candidate. It is then crucial that the scalar potential has a global minimum and it corresponds to the desired $(0,0,v)$ configuration. We revisit the formulation of the scalar potential leading to the theoretical conditions for consistency. 

We perform a comprehensive scan of the $\Z3$ 3HDM parameter space. Every relevant theoretical, experimental and astrophysical constraint is applied. This includes the recent and stringent direct detection bounds by the LZ collaboration~\cite{LZ:2024zvo} and the collider data on the $125\,\textrm{GeV}$ Higgs boson invisible branching ratio from ATLAS~\cite{ATLAS:2023tkt}. The validation process includes comparison of the one property of DM that is precisely measured with the numerically calculated prediction, the DM relic density of
$\Omega_{\textrm{DM}}= 0.1200 \pm 0.0012$~\cite{Planck:2018vyg}. Given the opposing tension in scalar DM models  between the measured relic abundance and the non-observation in detection experiments~\cite{Goncalves:2025snm}, we follow a Machine Learning (ML) approach to perform the constrained sampling of the model.

In the past 10 years, ML has been integrated in many of the different areas of Particle Physics, with modern reviews documenting the constant advancement of applications~\cite{Feickert:2021ajf,Shanahan:2022ifi,Plehn:2022ftl,Duarte:2025qbk}. 
Our goal is to efficiently obtain solutions that meet a long list of constraints, as a optimization problem. There have been multiple cases of success with different approaches, with their benefits and disadvantages~\cite{Caron:2016hib,Caron:2019xkx,Staub:2019xhl,Kronheim:2020vct,Hollingsworth:2021sii,Goodsell:2022beo,Diaz:2024yfu,Diaz:2024sxg,AbdusSalam:2024obf,Batra:2025amk,Hammad:2025wst,Biekotter:2025gkp,Duarte:2025qbk}. Starting from a model where valid training data is not easily obtainable, we 
follow a variant of the Evolutionary Strategy algorithm, first introduced in Ref.~\cite{deSouza:2022uhk}, 
combined with an histogram based anomaly detection procedure used for Novelty
Reward, developed in Ref.~\cite{Romao:2024gjx}. We improve on the previous methodology by considering a novel prototype selection approach, using information of already completed runs to optimally select the starting points for future runs, in regions that are less populated. The technique presented allows the automatized identification of representative benchmark solutions of a chosen subset in parameter and/or observable space. The approach also enables more efficient exploration of the model, yielding a wider range of allowed dark matter masses and expanding upon the existing literature \cite{Aranda:2019vda}. 

The work is organized as follows.
In Section~\ref{sec:model},  we recapitulate the $\Z3$ symmetric scalar potential and discuss the conditions for the boundedness from below and the desired $(0,0,v)$ vacua to be the global minima. In Section~\ref{sec:strategy}, we describe the strategy developed for sampling the valid parameter space of the model. In Section~\ref{sec:constraints}, we list the constraints included in the ML sampling algorithm. Sections~\ref{sec:results} is dedicated to our results, first by setting the mixing angle $\theta$ to $\tfrac{\pi}{4}$, under the assumptions of Ref.~\cite{Aranda:2019vda}, followed by a study on the alternative solution to the vertex $ZH_iA_i \sim \cos(2\theta)$ vanishing. After summarizing our conclusions in Section~\ref{sec:conclusions}, we present additional details regarding contributions to the relic density in Appendix~\ref{app:appendixA} and methods to improve our sampling in Appendix~\ref{app:appendixB}.

\section{The \texorpdfstring{\boldmath$\Zbold3$}{Z3} 3HDM}\label{sec:model}

\subsection{The Model}

We consider a 3HDM with a Lagrangian invariant under a $\Z3$ symmetry, satisfying
\begin{align}
    \label{eq:Z3symmetry}
    \Z3 :&\ \phi_1 \to \omega\,\phi_1\,,\quad \phi_2 \to \omega^2\,\phi_2\,,\quad
    \phi_3\to  \phi_3\, ,
\end{align}
with $\omega = e^{\tfrac{2\pi i}{3}}$. With the parametrization used in Refs.~\cite{Das:2019yad,Boto:2021qgu,Boto:2023nyi,Romao:2024gjx}, the resulting potential takes the form 
\begin{align}
    V = \, & m^2_{11}(\phi_1^\dagger \phi_1) 
    + m^2_{22}(\phi_2^\dagger \phi_2) 
    + m^2_{33}(\phi_3^\dagger \phi_3) \nonumber \\[4pt]
    & + \lambda_1(\phi_1^\dagger \phi_1)^2 
    + \lambda_2(\phi_2^\dagger \phi_2)^2 
    + \lambda_3(\phi_3^\dagger \phi_3)^2 \nonumber \\[4pt]
    & + \lambda_4(\phi_1^\dagger \phi_1)(\phi_2^\dagger \phi_2) 
    + \lambda_5(\phi_1^\dagger \phi_1)(\phi_3^\dagger \phi_3) 
    + \lambda_6(\phi_2^\dagger \phi_2)(\phi_3^\dagger \phi_3) \nonumber \\[4pt]
    & + \lambda_7(\phi_1^\dagger \phi_2)(\phi_2^\dagger \phi_1)
    + \lambda_8(\phi_1^\dagger \phi_3)(\phi_3^\dagger \phi_1) 
    + \lambda_9(\phi_2^\dagger \phi_3)(\phi_3^\dagger \phi_2) \nonumber \\[4pt]
    & + \Big[ \lambda_{10}(\phi_1^\dagger \phi_2)(\phi_1^\dagger \phi_3) 
    + \lambda_{11}(\phi_1^\dagger \phi_2)(\phi_3^\dagger \phi_2)
    + \lambda_{12}(\phi_1^\dagger \phi_3)(\phi_2^\dagger \phi_3) + \textrm{ h.c.} \Big] \, , \label{eq:Z3potential}
\end{align}
having the direct relation to the convention used in Ref.~\cite{Aranda:2019vda} by,
\begin{align}
    \lambda_1 &\rightarrow \lambda_{11} \, , \quad \lambda_2 \rightarrow \lambda_{22} \, , \quad \lambda_3 \rightarrow \lambda_{33} \, , \quad \lambda_4 \rightarrow \lambda_{12} \, , \nonumber \\
    \lambda_5 &\rightarrow \lambda_{13} \, , \quad \lambda_6 \rightarrow \lambda_{23} \, , \quad 
    \lambda_7 \rightarrow \lambda_{12}'\, , \quad \lambda_8 \rightarrow \lambda_{13}'\, ,  \nonumber \\ \lambda_9 &\rightarrow \lambda_{23}'\, , \quad
    \lambda_{10} \rightarrow \lambda_{1} \, , \quad \lambda_{11} \rightarrow \lambda_{2} \, , \quad \lambda_{12} \rightarrow \lambda_{3} \, .
\end{align}
Aiming for a scenario with two DM states, we consider the vacuum alignment $\langle \phi_1 \rangle = \langle \phi_2 \rangle =0 $ and $\langle \phi_3 \rangle = v$, thus keeping the $\Z3$ symmetry unbroken, defined by the transformation in Eq.~\eqref{eq:Z3symmetry}. After Spontaneous
Symmetry Breaking (SSB), the doublets are developed in the symmetry
basis as
\begin{equation}
\phi_1 = 
\begin{pmatrix}
H_1^+ \\[6pt]
\dfrac{x_1 + i y_1}{\sqrt{2}}
\end{pmatrix},
\qquad
\phi_2 = 
\begin{pmatrix}
H_2^+ \\[6pt]
\dfrac{x_2 + i y_2}{\sqrt{2}}
\end{pmatrix},
\qquad
\phi_3 = 
\begin{pmatrix}
G^+ \\[6pt]
\dfrac{v + h + i G^0}{\sqrt{2}}
\end{pmatrix}.
\end{equation}
The corresponding minimization condition is given by
\begin{equation}\label{eq:minimization}
    v^2 = \frac{m^2_h}{2 \lambda_3} \, .
\end{equation}
The mass matrices of the fields $x_i$ and $y_i$ in the electroweak basis are, for the neutral CP-even scalars,
\begin{equation}
    \mathcal{M}_{\textrm{even}}^2 =
    \begin{pmatrix}
    m_{11}^2 + \dfrac{1}{2}v^2(\lambda_{5}+\lambda_{8}) & \dfrac{1}{2}v^2\lambda_{12} & 0 \\[6pt]
    \dfrac{1}{2}v^2\lambda_{12} & m_{22}^2 + \dfrac{1}{2}v^2(\lambda_{6}+\lambda_{9}) & 0 \\[6pt]
    0 & 0 & 2\lambda_3 v^2
\end{pmatrix}
\end{equation}
for the CP-odd scalars, 
\begin{equation}
\mathcal{M}_{\textrm{odd}}^2 =
\begin{pmatrix}
m_{11}^2 + \dfrac{1}{2}v^2(\lambda_{5}+\lambda_{8}) & -\dfrac{1}{2}v^2\lambda_{12} & 0 \\[6pt]
-\dfrac{1}{2}v^2\lambda_{12} & m_{22}^2 + \dfrac{1}{2}v^2(\lambda_{6}+\lambda_{9}) & 0 \\[6pt]
0 & 0 & 0
\end{pmatrix}
\end{equation}
and for the charged
\begin{equation}
\mathcal{M}_{\textrm{charged}}^2 =
\begin{pmatrix}
m_{11}^2 + \dfrac{1}{2}v^2\lambda_{5} & 0 & 0 \\[6pt]
0 & m_{22}^2 + \dfrac{1}{2}v^2\lambda_{6} & 0 \\[6pt]
0 & 0 & 0
\end{pmatrix} \, .
\end{equation}
which is already diagonal.
The masses in the physical basis for the neutral scalars can then be obtained by applying the rotation 
\begin{equation}
    R \mathcal{M}^2R^T = \textrm{diag}(m^2_1 \, , m^2_2 \, , m^2_3)
\end{equation}
with a matrix of the form
\begin{equation}
R_{\theta_i} = 
\begin{pmatrix}
\cos\theta_i & \sin\theta_i & 0 \\[4pt]
-\sin\theta_i & \cos\theta_i & 0 \\
0 & 0 & 1
\end{pmatrix} \, .
\end{equation}
Here, $\theta_i = (\theta_x \, , \theta_y)$ for the scalar fields $x_i$ and the pseudoscalar fields $y_i$, respectively, with $-\pi/2 < \theta_i \leq \pi/2$. The diagonalization yields that $\tan(2\theta_x) = - \tan(2\theta_y)$, resulting in a simpler transformation that depends only on a single angle $\theta$, of the form
\begin{equation}
    \label{eq:rotation}
    \left( 
    \begin{array}{c}
    H_1 \\
    H_2
    \end{array}
    \right)
    =
    \left( 
    \begin{array}{cc}
    c_{\theta} & s_{\theta} \\
    -s_{\theta} & c_{\theta}
    \end{array}
    \right) 
    \left( 
    \begin{array}{c}
    x_1 \\
    x_2
    \end{array}
    \right) \, , \quad 
    \left( 
    \begin{array}{c}
    A_1 \\
    A_2
    \end{array}
    \right)
    =
    \left( 
    \begin{array}{cc}
    c_{\theta} & -s_{\theta} \\
    s_{\theta} & c_{\theta}
    \end{array}
    \right) 
    \left( 
    \begin{array}{c}
    y_1 \\
    y_2
    \end{array}
    \right) \, ,
\end{equation}
with $c_{\theta} = \cos(\theta)$ and $s_{\theta} = \sin(\theta)$. The charged-scalar fields require no further transformations as they are already in the physical basis. The resulting expressions for the masses are then,
\begin{align}
m^2_{H_1} &= m^2_{A_1} = 
\left(m_{11}^2 + \frac{1}{2}v^2(\lambda_5 + \lambda_8)\right)c^2_{\theta}
+ \left(m_{22}^2 + \frac{1}{2}v^2(\lambda_6 + \lambda_9)\right)s^2_{\theta}
 + \lambda_{12} v^2 s_{\theta} c_{\theta}\, , \nonumber \\[4pt]
m^2_{H_2} &= m^2_{A_2} =
\left(m_{22}^2 + \frac{1}{2}v^2(\lambda_6 + \lambda_9)\right)c^2_{\theta}
+ \left(m_{11}^2 + \frac{1}{2}v^2(\lambda_5 + \lambda_8)\right)s^2_{\theta}
 - \lambda_{12} v^2 s_{\theta} c_{\theta}\, , \nonumber \\[4pt]
m^2_{H^{\pm}_1} &= m_{11}^2 + \frac{\lambda_5 v^2}{2}\, , \nonumber \\[4pt]
m^2_{H^{\pm}_2} &= m_{22}^2 + \frac{\lambda_6 v^2}{2}\, .
\end{align}

 Following Ref.~\cite{Aranda:2019vda}, we take as independent parameters
\begin{equation}
    m_{H_1} \, , \, m_{H_2} \, , \, m_{H^{\pm}_1} \, , \, m_{H^{\pm}_2}\, , \, \theta \, , \, g_1 \, , \, g_2 \, , \, \lambda_1 \, , \, \lambda_4 \, , \, \lambda_7 \, , \, \lambda_{10} \, , \,\lambda_{11} \, , \,
    \label{eq:independentparameters}
\end{equation}
where $\theta$ is the mixing angle between $\phi_1$ and $\phi_2$, $g_1 = g_{h H_1 H_1}/v$ is the couplings between $H_1$ and the $125\,\textrm{GeV}$ Higgs and $g_2 = g_{h H_1 H_2}/v$. In this work, the signs of $g_1$ and $g_2$ are defined by the Feynman rules and are symmetric of the ones in Ref.~\cite{Aranda:2019vda}, meaning $g_i \rightarrow -g_i$ in our analysis. We know the mass of the $h_{125}$ and the vev to be 
\begin{equation}
v = \left(\sqrt{2} G_F\right)^{-1/2} \approx 246 \textrm{ GeV}\, , \quad  m_h = 125 \textrm{ GeV}\, ,
\end{equation}
from which we can get $m^2_{33}$ and $\lambda_3$ with the stationarity conditions in Eq.~\eqref{eq:minimization}. The other parameters of the potential can be obtained from the above. We detect a mismatch between the expressions of Ref.~\cite{Aranda:2019vda} and the ones we derived, coinciding only for $\cos(2\theta)=0$,
\begin{align}
    \lambda_{3} &= \frac{m_{h}^{2}}{2 v^{2}} \nonumber \, , \\[4pt]
    \lambda_{5} &= -g_{1} + \frac{2 m_{H_{1}^{\pm}}^{2}}{v^{2}} - \frac{2 m_{H_{1}^{0}}^{2}}{v^{2}} + g_{2} \tan(\theta) \nonumber \, ,\\[4pt]
    \lambda_{6} &= -g_{1} + \frac{2 m_{H_{2}^{\pm}}^{2}}{v^{2}} - \frac{2 m_{H_{1}^{0}}^{2}}{v^{2}} - g_{2} \cot(\theta) \nonumber \, ,\\[4pt]
    \lambda_{8} &= \frac{1}{v^{2}} \Big( -2 m_{H_{1}^{\pm}}^{2} + m_{H_{1}^{0}}^{2} + m_{H_{2}^{0}}^{2} + \big( m_{H_{1}^{0}}^{2} - m_{H_{2}^{0}}^{2} \big)\cos(2\theta) \Big) \nonumber \, ,\\[4pt]
    \lambda_{9} &= \frac{1}{v^{2}} \Big( -2 m_{H_{2}^{\pm}}^{2} + m_{H_{1}^{0}}^{2} + m_{H_{2}^{0}}^{2} + \big( -m_{H_{1}^{0}}^{2} + m_{H_{2}^{0}}^{2} \big)\cos(2\theta) \Big) \nonumber \, ,\\[4pt]
    \lambda_{12} &= \frac{1}{v^{2}} \Big( (m_{H_{1}^{0}}^{2} - m_{H_{2}^{0}}^{2}) \sin(2\theta) \Big) \nonumber \, ,\\[4pt]
    m_{11}^{2} &= m_{H_{1}^{0}}^{2} + \frac{1}{2} v^{2} \Big( g_{1} -  g_{2}  \tan(\theta) \Big) \nonumber \, ,\\[4pt]
    m_{22}^{2} &=  m_{H_{2}^{0}}^{2} + \frac{1}{2}v^{2} \Big( g_{1}  + g_{2} \cot(\theta) \Big) \nonumber \, ,\\[4pt]
    m_{33}^{2} &= -\lambda_{3} v^{2} \, .
\end{align}
As the squared masses are all positive, we may conclude that we are at the local minimum. Moreover, we obtain the mass degeneracy conditions $m_{H_1} = m_{A_1}$ and $m_{H_2} = m_{A_2}$. This relation is not arbitrary, as it arises directly from the symmetry and the vacuum configuration. Therefore, we use $m_{H_1} = m_{A_1} = m_{\textrm{DM}}$ henceforth, for the analysis and in the following plots.

Regarding the interactions with fermions and gauge bosons, all SM fields are chosen to be even under $\Z3$. Therefore, all fermion fields couple only with $\phi_3$ and the Yukawa couplings follow as identical to the SM. This realises a Type-I model which has no FCNCs at tree-level. 

\subsection{Bounded from below conditions}

As a consistency requirement for any physical theory, the scalar potential must satisfy conditions
that ensure it possesses a stable minimum. The potential given in Eq.~\eqref{eq:Z3potential} must be bounded from below, meaning that there is no direction in field space along which the value of the potential tends to minus infinity. Such directions can be classified into neutral directions, which do not break electric charge, or charge-breaking (CB) directions. Ref.~\cite{Faro:2019vcd} has shown, for the specific case of the $U(1)\times U(1)$ symmetric 3HDM, that the potential can be
unbounded from below along CB directions despite satisfying the conditions along neutral directions.

Following the procedure in Ref.~\cite{Boto:2022uwv},
the quartic part of the potential can be rewritten into a piece invariant under rephasings, similar for all symmetry constrained 3HDM potentials, with terms related to neutral ($V_N$) and CB ($V_{CB}$) directions, and the $\Z3$-specific terms as
\begin{equation}
    V_{\text{quartic}} = V_N + V_{CB} + V_{\Z3} \, ,
\end{equation}
where
\begin{align}
    V_N &= \lambda_1(\phi_1^\dagger \phi_1)^2 + \lambda_2(\phi_2^\dagger \phi_2)^2 + \lambda_3(\phi_3^\dagger \phi_3)^2 \nonumber \\[4pt]
    & \quad + (\lambda_4 + \lambda_7)(\phi_1^\dagger \phi_1)(\phi_2^\dagger \phi_2) 
    + (\lambda_5 + \lambda_8)(\phi_1^\dagger \phi_1)(\phi_3^\dagger \phi_3)
    + (\lambda_6 + \lambda_9)(\phi_2^\dagger \phi_2)(\phi_3^\dagger \phi_3) \,, \\[8pt]
    V_{CB} &= - \lambda_7 z_{12} - \lambda_8 z_{13} - \lambda_9 z_{23}\,, \\[8pt]
    V_{\Z3} &= \lambda_{10}(\phi_1^\dagger \phi_2)(\phi_1^\dagger \phi_3) 
    + \lambda_{11}(\phi_1^\dagger \phi_2)(\phi_3^\dagger \phi_2)
    + \lambda_{12}(\phi_1^\dagger \phi_3)(\phi_2^\dagger \phi_3) + \textrm{ h.c.} \, ,
\end{align}
with the definition~\cite{Faro:2019vcd}
\begin{equation}
  \label{eq:25}
  z_{ij}= (\phi_i^\dagger\phi_i) (\phi_j^\dagger\phi_j)
  - (\phi_i^\dagger\phi_j) (\phi_j^\dagger\phi_i) \quad \text{(no sum)}\, .
\end{equation}
The vevs of the doublets are parametrized as
\begin{equation}
    \label{eq:parBFB}
    \phi_1 = \sqrt{r_1} 
    \begin{pmatrix}
        \sin(\alpha_1) \\
        \cos(\alpha_1) e^{i\beta_1}
    \end{pmatrix} \, , \quad
    \phi_2 = \sqrt{r_2} e^{i\gamma} 
    \begin{pmatrix}
        \sin(\alpha_2) \\
        \cos(\alpha_2) e^{i\beta_2}
    \end{pmatrix} \, , \quad
    \phi_3 = \sqrt{r_3} 
    \begin{pmatrix}
        0 \\ 1
    \end{pmatrix} \, ,
\end{equation}
allowing $V_N$ to be written in the form 
\begin{equation}
  \label{eq:neutralquarticform}
  V_N= \frac{1}{2} \sum_{ij} r_i A_{ij} r_j \,,\quad\text{with}\quad
  (A)_{ij}=
    \begin{pmatrix}
    2\lambda_{1}&\lambda_{4}+\lambda_{7}&\lambda_{5}+\lambda_{8}\\
    \lambda_{4}+\lambda_{7}&2\lambda_{2}&\lambda_{6}+\lambda_{9}\\
    \lambda_{5}+\lambda_{8}&\lambda_{6}+\lambda_{9}&2\lambda_{3}
  \end{pmatrix}  \, ,
\end{equation} 
which, following Klimenko~\cite{Klimenko:1984qx} and Kannike~\cite{Kannike:2012pe}, allows the formulation of the neutral copositivity conditions that assure this part of the potential is BFB, by the quadratic form $A$ being positive definite for $r_i \geq 0$. 
The full $\Z3$ symmetric potential, however, has no known necessary and sufficient conditions, so we follow the method similar to Ref.~\cite{Boto:2022uwv} to, at least, obtain sufficient conditions by bounding the potential with a lower one that can be reduced to a quadratic form, as in Eq.~\eqref{eq:neutralquarticform}. The CB part can be simplified to
\begin{equation}
    V_{CB} \geq V^{\textrm{lower}}_{CB} = r_1 r_2 \textrm{min}(0,-\lambda_7) + r_1 r_3 \textrm{min}(0,-\lambda_8) + r_2 r_3\textrm{min}(0,-\lambda_9) \, ,
\end{equation}
and the $\Z3$-specific part yields
\begin{equation}
    V_{Z_3} \geq V^{\textrm{lower}}_{Z_3} = -2(|\lambda_{10}|+|\lambda_{11}|) r_1 r_2
    -2(|\lambda_{10}|+|\lambda_{12}|) r_1 r_3
    -2(|\lambda_{11}|+|\lambda_{12}|) r_2 r_3 \, .
\end{equation}
Finally, the sufficient conditions for the $\Z3$ potential to be BFB consist of the requirement that the quadratic form, $A$ in Eq.~\eqref{eq:neutralquarticform}, is positive definite with the following entries $A_{ij}$~\cite{Romao:2024gjx,Boto:2025raf},
\begin{align}\label{eq:bfbfull}
    A_{11} &= 2\lambda_1\, , \nonumber \\
    A_{12} &= \lambda_4 + \lambda_7 + \min(0,-\lambda_7) - 2|\lambda_{10}| - 2|\lambda_{11}|\, , \nonumber \\
    A_{13} &= \lambda_5 + \lambda_8 + \min(0,-\lambda_8) - 2|\lambda_{10}| - 2|\lambda_{12}|\, , \nonumber \\
    A_{22} &= 2\lambda_2\, , \nonumber \\
    A_{23} &= \lambda_6 + \lambda_9 + \min(0,-\lambda_9) - 2|\lambda_{11}| - 2|\lambda_{12}|\, , \nonumber \\
    A_{33} &= 2\lambda_3\, .
\end{align}
The positivity conditions that the complete BFB matrix $A_{ij}$ has to obey are
\begin{align}
&A_{11} \geq 0\, ,\qquad A_{22} \geq 0\, ,\qquad A_{33} \geq 0\, , \\[2mm]
&\bar{A}_{12} = \sqrt{A_{11}A_{22}} + A_{12} \geq 0\, ,\qquad
\bar{A}_{13} = \sqrt{A_{11}A_{33}} + A_{13} \geq 0\, ,\qquad
\bar{A}_{23} = \sqrt{A_{22}A_{33}} + A_{23} \geq 0 \, , \\[2mm]
&\sqrt{A_{11}A_{22}A_{33}}
+ A_{12}\sqrt{A_{33}}
+ A_{13}\sqrt{A_{22}}
+ A_{23}\sqrt{A_{11}}
+ \sqrt{2\bar{A}_{12}\bar{A}_{13}\bar{A}_{23}}
\geq 0 \, . \label{eq:BFBAaux}
\end{align}

\subsection{Beyond Sufficient Conditions}
\label{subsec:beyond}

The conditions of Eq.~\eqref{eq:bfbfull} are sufficient, but not necessary: a point may fail them and still support no runaway direction, i.e., there exists no direction in field space along which the potential decreases without bound. In this case, the scan would silently discard a physically valid point, potentially missing important physics.
We check the global boundedness from bellow hierarchically: points passing Eq.~\eqref{eq:bfbfull} are accepted as BFB at negligible cost, and only points failing them undergo an explicit numerical search for an unbounded direction of the potential.

Boundedness can be assessed entirely by the quartic terms. Rescaling all fields, $\phi_i \to \lambda\,\phi_i$, the potential separates according to its degree in the fields,
\begin{equation}
    V(\lambda) = \lambda^4\, V_4(\hat r, \varphi) + \lambda^2\, V_2(\hat r, \varphi)\, ,
    \label{eq:quarticdecomposition}
\end{equation}
where $V_4$ collects the quartic couplings; $V_2$ the quadratic terms, composed  of soft masses; $\hat r$ denotes the direction in field space (the ratios of the $r_i$ of Eq.~\eqref{eq:parBFB}), and $\varphi = (\alpha_1, \beta_1, \alpha_2, \beta_2, \gamma)$ the five angles. As $\lambda \to \infty$, the quartic term dominates, so the potential is BFB if and only if
\begin{equation}
    V_4^{\mathrm{min}} = \min_{\substack{(r_1, r_2, r_3) \in \Delta \\ \varphi \in [-\pi, \pi]^5}} V_4 \;\geq\; 0\, ,
    \label{eq:V4min}
\end{equation}
with the magnitudes restricted to the compact simplex $\Delta = \{r_1 + r_2 + r_3 = 1,\ r_i \geq 0\}$, since $V_4$ is homogeneous of degree two in the $r_i$ and each direction is then represented exactly once.

The minimization of Eq.~\eqref{eq:V4min} is numerically tractable. We evaluate $V_4$ on a Sobol low-discrepancy sequence~\cite{sobol1967distribution}, whose quasi-random coverage of the domain efficiently identifies the candidate runaway directions, and then minimize locally from the lowest-valued samples, which converges rapidly. A sampled direction with $V_4 < 0$ immediately certifies unboundedness, whereas a positive $V_4^{\mathrm{min}}$ is only as reliable as the coverage of the search. We validated the method against a full global optimization on a subsample of points and, finding agreement, employ it alone in the scans. In the scan, the analytical conditions remain the fast filter, with the numerical check triggered only upon their failure. An extra offline check using \texttt{Minuit} was performed, validating both the asymptotic and the finite behaviour of the potential on a subset of points that failed the sufficient conditions but were still found to be BFB by the numerical check.

\subsection{Global Minimum}

Besides imposing the potential be BFB, we must guarantee the minimum we want to study is also the global minimum of the potential.
We are interested in the \texttt{2-Inert} minimum given by the vacuum structure $(0,0,v)$, as it allows for a multi-component DM scenario. To check if we are at the global minimum, a routine was implemented in the minimization program \texttt{Minuit} \cite{James:1975dr} by CERN. With this procedure, after imposing the BFB conditions, we found two other competing minima, $(v,0,0)$ and $(0,v,0)$. These can be written in terms of the potential parameters on that point as 
\begin{align}
    &\text{The case } v_1 = v \, , v_2 = 0 \, , v_3 = 0 \textrm{ : } \quad V^{\textrm{min}}_{(v,0,0)} = - \frac{m^4_{11}}{4\lambda_1} \, , \\
    &\text{The case } v_1 = 0 \, , v_2 = v \, , v_3 = 0 \textrm{ : } \quad V^{\textrm{min}}_{(0,v,0)} = - \frac{m^4_{22}}{4\lambda_2} \, , \\
    \label{eq:Z3minima}
    &\text{The 2-Inert vacua } v_1 = v_2 = 0\, , v_3 = v \textrm{ : } \quad V^{\textrm{min}}_{(0,0,v)} = - \frac{m^4_{33}}{4\lambda_3} \, .
\end{align} 
Given these conditions, for a given point, we may compare the value of the potential with these two cases and discard the point if either of the minima falls below the desired 2-Inert case in Eq.~\eqref{eq:Z3minima}. Then, as a sanity check, we confirm numerically that indeed no lower-lying minimum is found.

\section{Machine Learning scan strategy}\label{sec:strategy}

We allow values for the parameters of Eq.~\eqref{eq:independentparameters} in the ranges
\begin{align}
&\
\lambda_1,\, \lambda_4,\, \lambda_7,\, \lambda_{10},\, \lambda_{11}  \in \pm\left[10^{-3},10\right] \, ;
\nonumber \\[8pt]
&
m_{H_1} = m_{A_1} \,, \, m_{H_2} = m_{A_2 }
\in \left[50,1000\right]\textrm{ GeV} \, ;
\nonumber \\[8pt]
&
m_{H_1^\pm},\,m_{H_2^\pm}\,
\in \left[70,1000\right] \textrm{ GeV} \, ; \nonumber \\[8pt]
&
g_1 \, , g_2 \, \in [-0.5,0.5] \, ; \quad
\theta \, \in \Big[  -\frac{\pi}{2},\frac{\pi}{2} \Big] \, ;
\label{eq:scanparameters}
\end{align}
and impose the condition, without loss of generality, that $m_{H_1} < m_{H_2}$. The lower limit on the mass of the charged scalars follows from Ref.~\cite{Pierce:2007ut} \footnote{We consider the reinterpretation of LEP-II chargino searches for inert scalars in Refs.~\cite{Pierce:2007ut, Lundstrom:2008ai,Belyaev:2016lok} as a conservative lower bound. Standard charged-Higgs searches, which assume fermionic decays, do not apply. Raising this bound to 100 GeV only removes the low-mass region with $m_{\rm DM} \lesssim 60$ GeV. A dedicated recast of LHC chargino searches to this scenario is left for future work.}. Although this bound has not been established within the context of the current model, we take it as a conservative lower bound on the masses of all charged scalars.

After rotating the DM fields to the mass basis, there are two distinct sectors: the dark-$\Z3$ sector, constituted by the fields in $\phi_1$ and $\phi_2$, and the active or SM sector, constituted by the fields in $\phi_3$ and all SM fermions. This differs from the two-component DM case, also with two additional inert doublet scalars of Ref.~\cite{Boto:2024tzp}, as it has three distinct sectors. Connections among different sectors can only occur due to gauge bosons or the interactions in the scalar potential.

\subsection{Evolutionary Strategy and Novelty Detection for Parameter Space Exploration}

We employ the ML-based scan strategy first described in Ref.~\cite{deSouza:2022uhk}. The astrophysical data from relic density, direct and indirect detection are considered in the optimisation procedure, taking advantage of its rapid convergence and capacity to explore the viable parameter space of a given model, as further demonstrated in Refs.~\cite{Romao:2024gjx,deSouza:2025uxb,deSouza:2025bpl,Boto:2025mmn,Boto:2025ovp,Boto:2026gzj}. The observables are calculated within a black-box, a combination of an in-house code and calls to public software, \texttt{HiggsTools}~\cite{Bahl:2022igd} and \texttt{micrOmegas}~\cite{Alguero:2023zol}, which are then introduced into the ML pipeline. Then, they are checked with a constraint function, $C$, defined as
\begin{equation}\label{eq:Cfunc}
	C(\mathcal{O}) = \max(0, -\mathcal{O} + \mathcal{O}_{LB}, \mathcal{O} - \mathcal{O}_{UB})\,,
\end{equation}
where $\mathcal{O}$ is the value of an observable or a constrained quantity, $\mathcal{O}_{LB}$ is its lower bound, and $\mathcal{O}_{UB}$ its upper bound. The constraint function $C(\mathcal{O})$ returns 0 if and only if $\mathcal{O}$ lies within the specified interval. Otherwise, it returns a positive number quantifying \textit{how far} the value of the observable lies from the specified bounds. Finding points that satisfy a given constraint is equivalent to minimizing the respective  $C(\mathcal{O})$ function. To account for multiple observables and parameter bounds simultaneously, one sums all the constraints $C(\mathcal{O})$ into a single loss function, $L$,\footnote{For a multi-objective alternative approach see Ref.~\cite{deSouza:2025uxb}.} as
\begin{equation} \label{eq:lossf}
	L(\hat{\theta}) = \sum_{i=1}^{N_c} C(\mathcal{O}_i(\hat{\theta})),
\end{equation}
where $\hat{\theta}$ is the input vector in the parameter space and $N_c$ is the number of constraints considered.  The sampling algorithm chosen to optimize the loss function in Eq.~\eqref{eq:lossf} is the Covariant Matrix Adaptation Evolutionary Strategy optimiser (CMA-ES)~\cite{CMAES-original,Hansen2001,hansen2023cmaevolutionstrategytutorial,nomura2024cmaessimplepractical}, later augmented with a novelty reward mechanism based on density estimation, using an Histogram Based Outlier Score (HBOS) \cite{HBOS,Kriegel} in Ref.~\cite{Romao:2024gjx}.
This procedure enhances exploration in some chosen parameters and/or observables of interest by adding a penalty $p$ to the loss function $L$.
The penalties are normalized to $p \in [0,1]$, such that new points receive a penalty of value 0, corresponding to a lower density, while points similar to previously explored ones receive penalties approaching maximum value of 1. This adjustment encourages CMA-ES to explore novel regions by penalising the loss function when constraints are satisfied and solutions are near already discovered areas. Throughout the analysis of the model in Section~\ref{sec:results}, we specify which parameters are receiving \textit{focus}, meaning that exploration in the plane of those quantities is actively being penalised and benefiting from the novelty reward features.

\subsection{Prototype Selection for Seed Acquisition} \label{subsec:Prototype}

Due to the exploitative and local search nature of CMA-ES, several independent runs of the algorithm are necessary to fully paint a picture of the global valid parameter space. This typically leads to convergence to a few disconnected regions, after some initial runs. The density of points in each of those regions is a good indicative of the easiness of convergence the algorithm has towards that particular region. This happens due to the shape of the constrained parameter space hypersurface that CMA-ES is learning. Being so, it is likely that a considerate portion of future runs will converge to similar previously discovered regions, thus hindering exploration. In order to obtain a global picture of the valid parameter space, one can force exploration by using seeds where new runs are to be initialised. These seeds can be selected amongst previously obtained points from sparsely populated regions, thus forcing CMA-ES, aided by the novelty reward mechanism, to effectively expand these regions.

The process of seed selection can be manually carried out by the user after careful data analysis to identify potential regions of interest for further exploration. Although this method can be valuable for seed selection immediately after initial runs, it can quickly turn into a time consuming and cumbersome process after a considerable amount of valid points has been amassed. 

In this work we develop three promising methods for seed selection based on known ML algorithms: Histogram Gradient Boosting Classifier (HGBC)~\cite{10.1007/978-3-030-37334-4_4}, Mini-Batch $k$-Means (MBKM)~\cite{10.1145/1772690.1772862}, and Local Outlier Factor (LOF)~\cite{LOF}. The prototype selection criteria for seeds is partially automated, while still allowing user intervention to customize acceptance quantiles, cluster selection, and other parameters. Further details on the implementation and how to employ such techniques can be found in Appendix~\ref{app:appendixB}. The three approaches are distinct and thus lead to different seed choices. This approach shares similarities with active learning, shown to have success in similar applications~\cite{Goodsell:2022beo}.

HGBC can be used to select seeds from globally underdense regions. After the addition of a mask of uniform noise over the data, we can train HGBC to distinguish the real data from the added noise. We can select the most outlying quantile in the collected valid points, typically $1\%$ or less, to be used as seeds. This way, we are selecting the points which are harder to distinguish from uniform noise, corresponding to the ones in less densely populated regions, in need for further exploration.

MBKM can be used to select seeds from the edges of clustered regions. Once trained on the dataset of valid points, we can select the points from each cluster that are farthest away from the respective centroid. This way, we are sampling seeds that lie on the frontier of each cluster. These seeds can then be used for future runs to effectively expand the frontiers of those clustered regions. $k$-Means has been previously successfully used for prototype selection for Quantum ML collider analysis~\cite{Peixoto:2022zzk}.

Lastly, we can use LOF to select seeds from the fringes of local neighbourhoods in the dataset of valid points. After training, LOF can give an anomaly score for each data point corresponding to its local density deviation with respect to its local neighbours. Like in the HGBC case, we can select a quantile ($ < 1\%$) for the highest anomaly scores in the data.

Moreover, there are regions which seem easy to explore when projected on a 2D plane, yet may be completely disconnected on the hypersurface supporting the valid points, given its non-convex and multi-dimensional nature. This creates dense shapes, with close neighbouring regions completely empty, as explicitly shown in Appendix~\ref{app:appendixB} where we further detail the new methodologies listed herein. As these areas are very populated, it becomes harder for HGBC to select points there without increasing the acceptance too much and end up with a sparse set of seeds, useless to focus on specific areas. To address this, one can use LOF to efficiently target the regions of interest as it can easily detect local density deviations and target the points that lie in the fringes of these areas. Alternatively, one can train MBKM, select the clusters of interest, isolate the points that lie furthest from its centroids and use them for seeds. This combined with a constraint that enforces exploration on a subset of the parameter space greatly improves the convergence rate. 

\section{Constraints}\label{sec:constraints}

For the single-objective optimization algorithm, we consider a loss function that sums all the different constraints in the form of $C(\mathcal{O})$ functions, defined in Eq.~\eqref{eq:Cfunc} for a chosen lower and upper bound. The constraint set incorporates theoretical bounds, experimental measurements, and selection cuts on an equal basis within the unified objective function.\footnote{There is however a possibility of imposing a hierarchical sequence~\cite{deSouza:2025uxb,deSouza:2025bpl}, especially important in scenarios where physical masses are derived quantities that must be positive. In this study we are not in this situation.} 
For the theoretical bounds, we apply the constraints from BFB and global minimum described in Section~\ref{sec:model}. Perturbative unitarity is considered following Refs.~\cite{Moretti:2015cwa,Bento:2022vsb} and compliance with the oblique parameters $\textrm{STU}$ from the global electroweak fit~\cite{Baak:2014ora}, calculating the numerical predictions using the formulae in Ref.~\cite{Grimus:2008nb}. 

As a Type-I model, all Yukawa fermion-Higgs couplings are identical to the SM. With the charged scalars also inert, the model bypasses all flavour constraints, such as bounds on masses from $B \to X_s \gamma$.

\subsection{Collider constraints}
 We calculate all processes at lowest non-trivial order and take all input parameters at the electroweak scale.
Our in-house program includes the latest LHC bounds on the $h_{125}$ signal strengths with the full Run~2 data collected at $13 \,\textrm{TeV}$, for the different production and decay modes,
following the ATLAS results in fig.$\,3$ of Ref.~\cite{ATLAS:2022vkf}. We require agreement with the discovered Higgs boson signal strengths at $2\sigma$. For a given production and decay channel, the Higgs signal strength is defined as 
\begin{equation}
\mu_i^f =
\frac{\sigma_i^{\mathrm{3HDM}}(pp\to h_{125}) }{\sigma_i^{\mathrm{SM}}(pp\to h_{125})} \times
\frac{\mathrm{BR}^{\mathrm{3HDM}}(h_{125}\to f)}{\mathrm{BR}^{\mathrm{SM}}(h_{125}\to f)}\, ,
	\label{eq:ss}
\end{equation}
The upper limit on the Higgs total decay width is set by Ref.~\cite{ParticleDataGroup:2024cfk} at $\Gamma_\textrm{tot}\leq 7.5\, \mathrm{MeV}$ and we forbid decays of SM gauge bosons into the new scalars, imposing~\cite{Hernandez-Sanchez:2020aop},
\begin{equation}
m_{H_i}+m_{H_i^\pm} \geq m_W^\pm\, , \hspace{2ex}
m_{A_i}+m_{H_i^\pm} \geq m_W^\pm\, , \hspace{2ex}
m_{H_i}+m_{A_i} \geq m_Z\, , \hspace{2ex}
2 m_{H_i^\pm}\geq m_Z \, .
\end{equation}
In order to evade the bounds from long-lived charged particle searches given in Ref.~\cite{Heisig:2018kfq}, we set the upper limit on the charged scalar lifetime of $\tau \leq 10^{-7}\textrm{s}$. 

If the DM mass $m_{\textrm{DM}} \lesssim m_{h_{125}}/2$, the Higgs boson can decay invisibly. This decay contributes to the Higgs total width and modifies its branching ratios. The invisible branching ratio at $95\%$ C.L. is given by
\begin{equation}
    \label{eq:BR_inv}
    \textrm{BR}(h_{125} \rightarrow \textrm{invisible}) = \frac{\Gamma(h_{125} \rightarrow \chi \chi)}{\Gamma^{\textrm{SM}}_{h_{125}} + \Gamma(h_{125} \rightarrow \chi \chi)} \leq 0.107 \pm 0.077 \quad (\textrm{95\% CL}) \, ,
\end{equation}
from a combined ATLAS analysis~\cite{ATLAS:2023tkt}, where $\Gamma(h_{125} \rightarrow \chi \chi)$ is the partial width of the Higgs into a pair of DM particles and the denominator is the total width of the Higgs boson in the given model.

The points are also passed through \texttt{HiggsTools-1.1.3}~\cite{Bahl:2022igd}, imposing current bounds from searches for additional scalars with the most recent module HiggsBounds (HB), version 1.6 of the dataset. We also require a $3 \sigma$ agreement with the $\Delta \chi^2$ test from the {HiggsSignals} module of \texttt{HiggsTools-1.1.3}. We later verified agreement with the new release of \texttt{HiggsTools-1.2} and the \texttt{HiggsBounds dataset-1.7}.

\subsection{Dark Matter Constraints}
\label{sec:DM_constraints}

We implement the model in \texttt{FeynMaster-2.1}~\cite{Fontes:2019wqh,Fontes:2021iue,Fontes:2025svw}, combining the public codes \texttt{FeynRules-2.3.49}~\cite{Christensen:2008py,Alloul:2013bka}, QGRAF 4.0.5~\cite{Nogueira:1991ex} and \texttt{FeynCalc-10.1.0}~\cite{Mertig:1990an,Shtabovenko:2016sxi,Shtabovenko:2020gxv}, that can then export the files necessary to run \texttt{micrOmegas-6.2.3}~\cite{Alguero:2023zol}, used for relic density, scattering amplitudes and annihilation cross section calculations.

\subsubsection{Relic Density} \label{subsec:RelicDensity}

The total DM relic density is given by the sum of the individual contributions from each DM component. We impose consistency with the measurement reported by the \textit{Planck} Collaboration~\cite{Planck:2018vyg} up to $3 \sigma$, 
\begin{equation}
    \Omega_T h^2 = \Omega_1 h^2 + \Omega_2 h^2 \, \overset{\textit{Planck}}{=} \, 0.1200 \pm 0.0012 .
\end{equation}
In order to achieve the correct results for the relic calculation in this model, one has to consider separate thermal histories for both doublets and employ the appropriate routine of \texttt{micrOMEGAs-6.2.3}, \texttt{DarkOmegaN}, as for higher temperatures, co-scattering of the form $\chi_i \;\textrm{SM} \rightarrow \chi_i' \; \textrm{SM}'$ and DM conversion processes are essential to maintain chemical equilibrium. A detailed discussion can be found in Appendix~\ref{app:appendixA}. Furthermore, as both DM components, $H_1$ and $A_1$, arise from the same doublet, they share the same couplings, resulting in an equal contribution from both components to the relic. The contribution from $H_2$ and $A_2$ is negligible, as they are heavier and decay completely.

\subsubsection{Direct Detection}

\begin{figure}[htbp!]
	\centering
	\includegraphics[width = 0.48\textwidth]{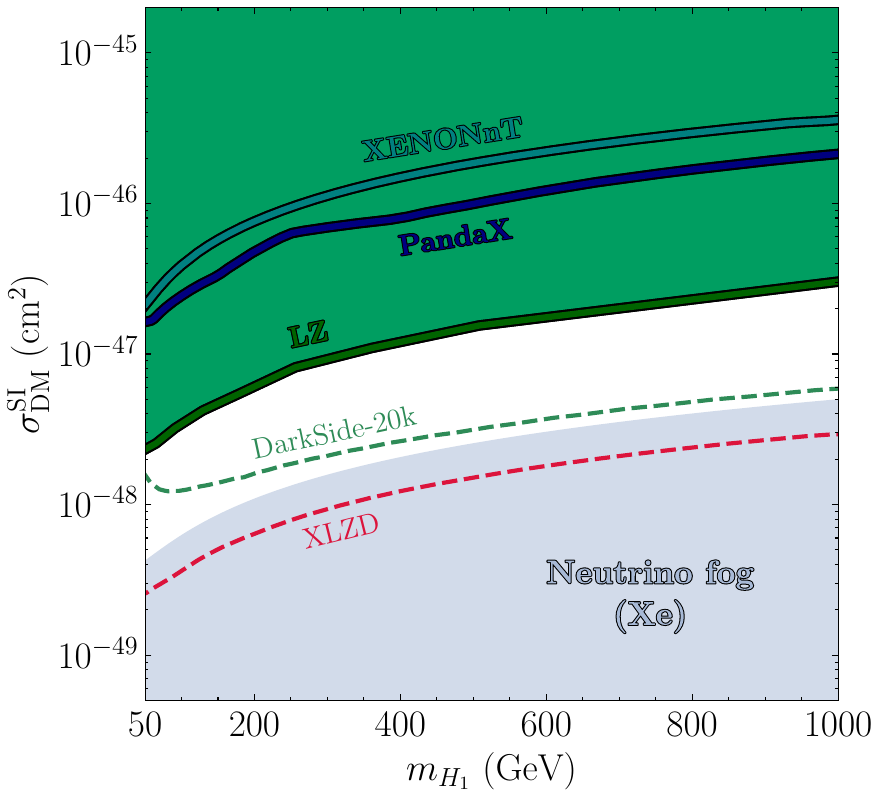}
	\caption[Literature benchmarks comparison]{In solid lines, the most recent direct detection limits from XENONnT~\cite{XENON:2025vwd}, PandaX-4T~\cite{PandaX:2024qfu} and LZ~\cite{LZ:2024zvo} for the mass region of interest. In dashed lines, the pojections for the upcoming DarkSide-20k~\cite{DarkSide-20k:2017zyg} and XLZD~\cite{XLZD:2024nsu} experiments, which is expected to breach the neutrino fog in the region of interest for the WIMP mass~\cite{OHare:2021utq}, shown as the grey shaded region. As a reminder, $m_{H_1}$ should read as $m_{H_1} = m_{A_1} = m_{\textrm{DM}}$. }
	\label{fig:DD-limits}
\end{figure}

The direct detection experiments look for the interaction of the DM particles with regular matter, through measuring the recoil energy of nuclei.
The most recent constraints from DM-nucleon scattering are provided by LZ 2025~\cite{LZ:2024zvo}, shown in Fig.~\ref{fig:DD-limits} along with the XLZD predictions~\cite{XLZD:2024nsu} and the lines for the neutrino fog region\footnote{The direct detection data files and general layout for all the related plots follows from the public repository by Ciaran O’Hare,
available in \url{https://github.com/cajohare/DirectDetectionPlots}.}~\cite{OHare:2021utq}.

To compare the calculated predictions with these limits, we follow the procedure described in Ref.~\cite{Belanger:2014bga}, computing the spin-independent (SI) DM-nucleus cross section for a Xenon target,
\begin{equation}
    \sigma_{\mathrm{SI}}^{\mathrm{Xe},k} =
    \frac{4 \mu_k^2}{\pi} 
    \frac{\left[ Z f_p + (A-Z) f_n \right]^2}{A^2} \, ,
\end{equation}
where $\mu_k$ is the reduced mass of the $k$-th DM component, and $f_p$ ($f_n$) denote the effective couplings to protons (neutrons).  
Since the dark sector comprises two DM components, we rescale the scattering cross section of each component according to its relative abundance,
\begin{equation}
    \sigma_{\mathrm{SI}}^{\mathrm{r},k} =
    \sigma_{\mathrm{SI}}^{\mathrm{Xe},k} \, \xi_k \, , \quad \xi_k = \frac{\Omega_k}{\Omega_T} \, .
\end{equation}

Besides regular direct detection constraints, we must also consider bounds that arise from inelastic scattering. Direct detection limits from CDMS-type experiments strongly constrain unsuppressed $Z$ mediated WIMP-nucleon scattering. In models where the $Z H_1 A_1$ vertex is non-zero, the process $H_1 N \rightarrow A_1 N$ 
is allowed unless the scalar-pseudoscalar mass splitting is such that the kinetic energy available in the halo DM is not enough to produce a signal~\cite{Ghosh:2021noq,Banerjee:2021oxc}, leading to the requirement
\begin{equation} \label{eq:mass-split}
 |m_{A_1}-m_{H_1}| \gtrsim \mathcal{O}(100~\mathrm{keV}) \, ,
\end{equation}
or, if the states are exactly degenerate, the coupling $Z H_1 A_1 \sim \cos(2\theta) $ must vanish. This motivates the choice of $\cos(2\theta) \rightarrow 0$, enforcing a safe limit in the minimal model, by construction.

\subsubsection{Indirect Detection}

Indirect detection consists of experimental bounds from the (non-)observation of annihilation or decay products of WIMPs in the Universe. The current limits from the Fermi-LAT satellite on photon fluxes from DM annihilation~\cite{Fermi-LAT:2015att} constrain $\langle \sigma v \rangle$ to be approximately $3\times 10^{-26}\,\mathrm{cm}^3\mathrm{/s}$ for light DM and up to $\sim 10^{-25}\,\mathrm{cm}^3\mathrm{/s}$ for heavier DM.

\begin{figure}[htbp!]
	\centering
	\includegraphics[width = 0.48\textwidth]{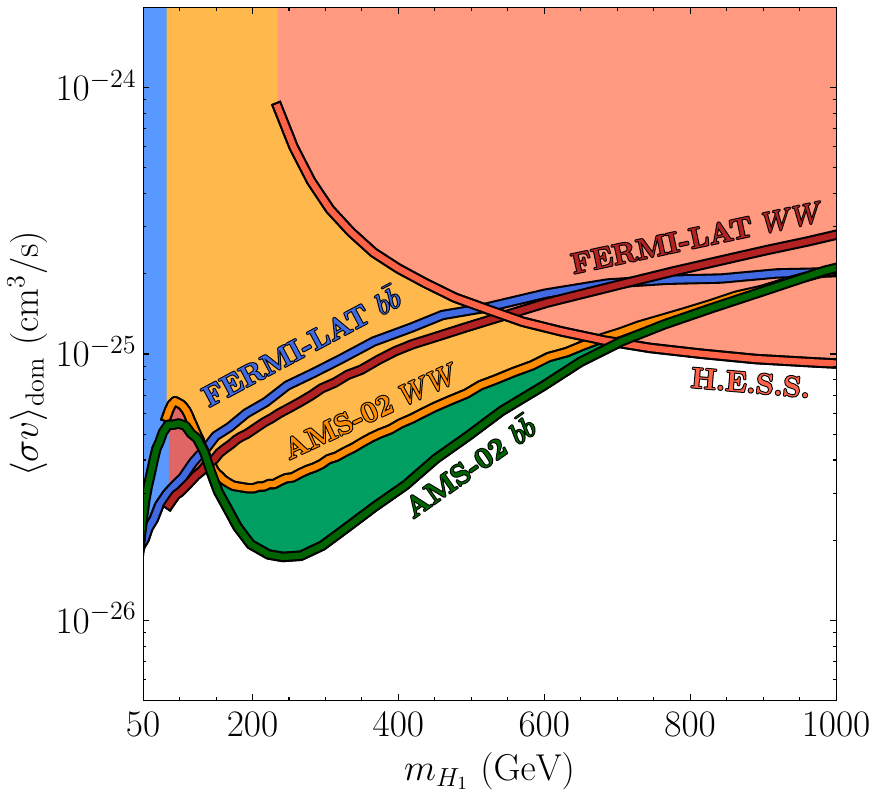}
	\caption[Literature benchmarks comparison]{Indirect detection constraints for $WW$ and $bb$ channels in the region of interest for the WIMP mass. The experimental lines originate from Fermi-LAT~\cite{Fermi-LAT:2015att}, H.E.S.S.~\cite{HESS:2022ygk} and AMS-02~\cite{AMS:2016oqu,AMS:2016brs}, following the approach in Ref.~\cite{Reinert:2017aga}.}
	\label{fig:ID-limits}
\end{figure}

Following Ref.~\cite{Belanger:2021lwd}, we find that the most relevant channels within the Fermi-LAT sensitivity region correspond to annihilations into $b\bar{b}$ and electroweak gauge bosons.  
We therefore require $ \langle \sigma v \rangle_{bb}$ and $ \langle \sigma v \rangle_{VV}$ to satisfy the $95\%$~C.L. Fermi-LAT bounds~\cite{Fermi-LAT:2015att}. Searches for antiprotons with AMS-02~\cite{AMS:2016oqu,AMS:2016brs} provide even more stringent constraints on WIMP DM. In this case, we again use $ \langle \sigma v \rangle_{bb}$ and $\langle \sigma v \rangle_{VV}$, and apply the bounds from Ref.~\cite{Reinert:2017aga}. Finally, we note that the H.E.S.S. telescope~\cite{HESS:2022ygk}, which observes gamma rays from the Galactic Center, sets the strongest constraints for DM masses above approximately $700~\mathrm{GeV}$. All these limits and their corresponding exclusion zones are displayed in Fig.~\ref{fig:ID-limits}.

To account for indirect detection limits, we follow the strategy of Ref.~\cite{Belanger:2021lwd} and use our \texttt{micrOMEGAs-6.2.3} model implementation to compute the thermally averaged annihilation cross section times velocity, $\langle \sigma v \rangle$. We performed an initial grid-search over the DM mass-relic plane, enforcing its compliance with Planck bounds on the relic density, and selected the dominant annihilation channels for each point. These constitute the $b\bar{b}$ channel for $m_{\textrm{DM}} \lesssim m_W$ and channels with $W$ and $Z$ boson contributions for $m_{\textrm{DM}} \gtrsim m_W \, , m_Z$. Due to the nature of experimental limits, which are given as separate lines for quarks pairs, such as $b\bar{b}$, and for gauge bosons as $VV$, we sum the contributions from annihilation into $WW$ and $ZZ$ as the same $\langle \sigma v \rangle_{VV}$ signal, while other channels, such as $b\bar{b}$ are considered separately. 

Considering separate thermal histories for both DM doublets, any possible rescaling of the indirect detection would not result in relevant changes, due to the negligible contribution to the relic of $m_{H_2}$ and $m_{A_2}$. They are heavier than $m_{H_1}$ and $m_{A_1}$ and decay completely. Therefore, in contrast to the direct detection limits with which one needs to be careful due to being extremely stringent, the results hold regardless of any possible rescaling of indirect measurements with the relic, as the contribution to the relic of the particles in the second thermal sector is negligible.

We leave the indirect detection constraint out of the primary pipeline and instead treat it as an additional validation step after sampling. Including the indirect constraint function $C(\theta)$ in the ML pipeline results in a significant increase in the time per generation, of order of five times (due to the long calculations of the annihilation and propagation of particles), while only superficially constraining the scans.  We observe the validity of this choice because we found, similarly to Ref.~\cite{Boto:2024tzp}, that for points already consistent with the relic abundance measurement the current indirect detection bounds result in only mild exclusions, of the order of $ \lesssim 1\%$, and therefore applying them only a posteriori allowed for considerably faster scans. 

\section{Results}
\label{sec:results}

Previous studies on the $\Z3$ 3HDM as a multi-component DM model have presented an analysis on the limit of $\theta = \frac{\pi}{4}$, as the condition guarantees the vertex $ZH_iA_i \sim \cos(2\theta)$ vanishes~\cite{Aranda:2019vda}. In this limit, the expressions derived for the model parameters coincide with ours and we are able to study the benchmark scenarios proposed. We start by reproducing the results obtained in Ref.~\cite{Aranda:2019vda} and point out the importance of applying the complete BFB conditions for the scalar potential, divided into the neutral, CB and $\Z3$-specific parts. 

We then continue in the limit $\theta = \frac{\pi}{4}$ without fixing the benchmark choices for the particle masses, taking into account all the necessary theoretical, collider and astrophysical constraints. Finally, we consider alternative solutions to $\cos(2\theta)=0$.

\subsection{The \texorpdfstring{\boldmath$\theta = \frac{\pi}{4}$}{theta} stability alignment}

\begin{figure}[h!]
	\centering
	\includegraphics[width = 0.98\textwidth]{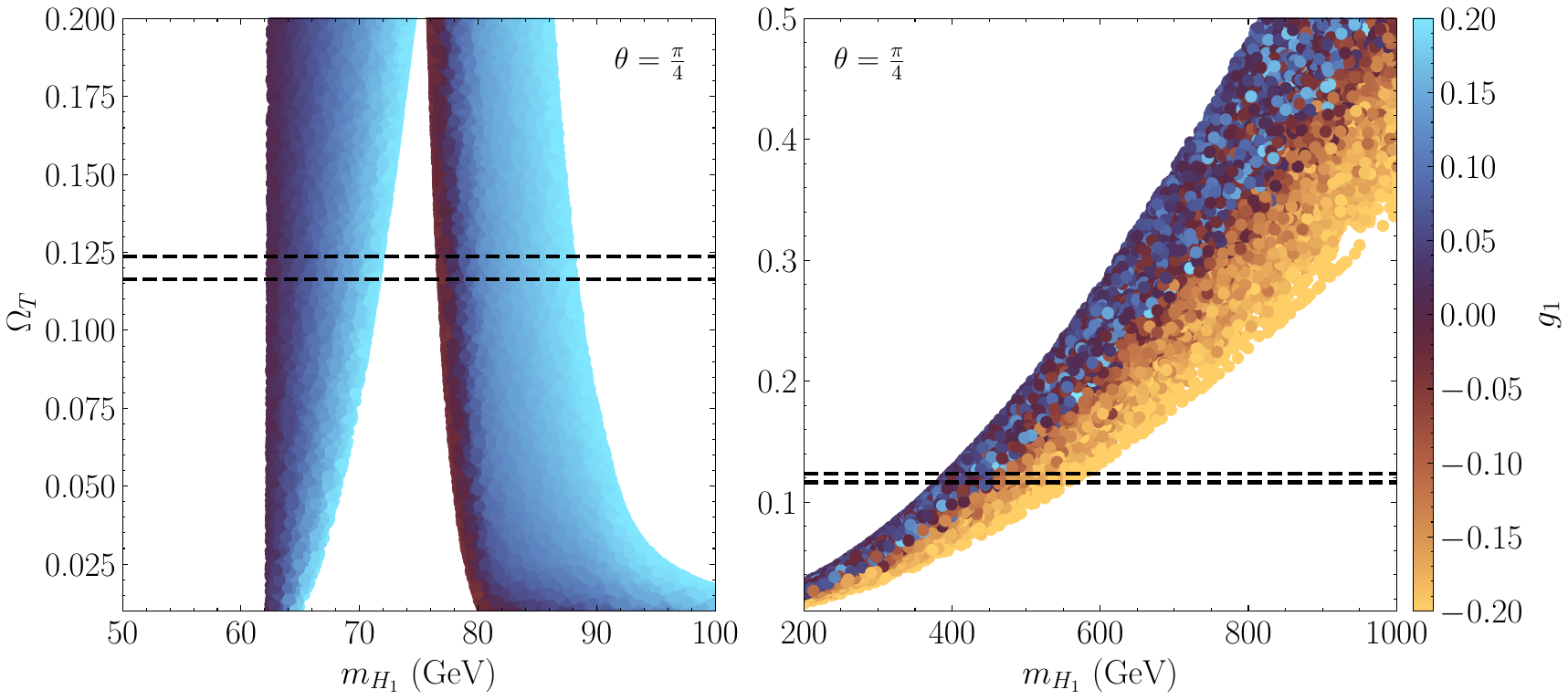}
	\captionsetup{font=small}
	\caption[Literature benchmarks comparison]{Benchmarks B and G from Ref.~\cite{Aranda:2019vda}, for the low and high mass regions, respectively, with colour code for $g_1$. Note that $g_1 \rightarrow -g_1$ when comparing plots with Ref.~\cite{Aranda:2019vda} due to the sign conventions. All the theoretical constraints are applied. \texttt{HiggsTools}, direct and indirect detection limits are also not enforced. The horizontal dashed lines are the bounds from relic density, considered at $3\sigma$.}
	\label{fig:Z3-NoCB-colorcode}
\end{figure}

Following Ref.~\cite{Aranda:2019vda}, benchmark scenario B consists of points in the low mass region, $45 \leq m_{\textrm{DM}} \leq 90\, \textrm{GeV}$ and benchmark G is the only scenario shown for the high mass region $400 \leq m_{\textrm{DM}} \leq 900\, \textrm{GeV}$. As described in Tab. II of Ref.~\cite{Aranda:2019vda}, both benchmarks fix a total of 8 quantities: 5 potential parameters to order $\mathcal{O}(0.1)$ and three relations of mass differences, while freely varying $\lambda_{11}$ $(\lambda_{11} \rightarrow \lambda_2)$ and the couplings, $g_1$ and $g_2$.
Taking as definitions for the mass difference variables the following relations, 
\begin{equation}\label{eq:massvariables}
\Delta_n = m_{H_2} - m_{H_1}\, , \quad
\Delta_c = m_{H_1^\pm} - m_{H_1}\, , \quad
\delta_c = m_{H_2^\pm} - m_{H_1^\pm}\, .
\end{equation}
With the introduced notation, benchmark B corresponds to fixing the variables,
\begin{equation}
\Delta_n = 50~\mathrm{GeV}\, , \quad
\Delta_c = 60~\mathrm{GeV}\, , \quad
\delta_c = 10~\mathrm{GeV}\, ,
\end{equation}
given that
\begin{equation}
m_{H_1} = m_{A_1} \ll m_{A_2} = m_{H_2} \ll m_{H_1^\pm} \sim m_{H_2^\pm} \, .
\end{equation}
Besides the mass-degenerate DM $H_1$ and $A_1$, all other inert particles are much heavier, not allowing co-annihilation with them. As the vertex $ZH_iA_i$ does not exist for the $\theta=\tfrac{\pi}{4}$ condition, there is also no co-annihilation between $H_1$ and $A_1$. Finally, channels such as $hH_1A_1$ are also forbidden as the model is CP-conserving. The only co-annihilation processes available are the ones that involve the coupling $g_1$, with the $H_1H_1h$ and $A_1A_1h$ vertices.

Benchmark G corresponds to fixing the variables as,
\begin{equation}
\Delta_n = 2~\mathrm{GeV}\, , \quad
\Delta_c = 0.8~\mathrm{GeV}\, , \quad
\delta_c = 0.5~\mathrm{GeV}\, ,
\end{equation}
with similar values between all neutral and between the charged scalars, 
\begin{equation}
m_{H_1} = m_{A_1} \sim m_{A_2} = m_{H_2} \sim m_{H_1^\pm} \sim m_{H_2^\pm} \, .
\end{equation}
In this scenario, new channels open up for DM co-annihilation through $h$ and $Z$ mediated processes.

 The plots of Fig.~\ref{fig:Z3-NoCB-colorcode} show the results of our scans for the two benchmarks scenarios B and G, with a colour code for the value of $g_1$. The constraints included are the neutral BFB conditions, STU, global minimum, unitarity, signal strengths, the refactored limits on DM searches, the DM lifetimes, the invisible decay and width of the SM-like Higgs. No agreement with DM direct detection, indirect detection or Higgs data  with \texttt{HiggsTools} is imposed. In these conditions, we are able to reproduce the behaviour of Figures 1 and 4 of Ref.~\cite{Aranda:2019vda}, with the exception of the points in the region below the $m_{\textrm{DM}} \lesssim m_h / 2 \, \textrm{ GeV}$, which we did not find in these scans given the latest bounds of LHC constraints on the invisible branching ratio of the Higgs. There is still a need to check the boundedness from below along other directions. A possible approach would be to check the sufficient conditions described. However, one would naively exclude these points using the sufficient conditions. Upon closer inspection on the runaway directions of the quartic terms of the potential, some of the points were found to be BFB and therefore viable. This poses a cautionary tale of not using sufficient conditions as exclusion criteria because they might mislead the study.

\begin{figure}[h!]
	\centering
	\includegraphics[width = 0.48\textwidth]{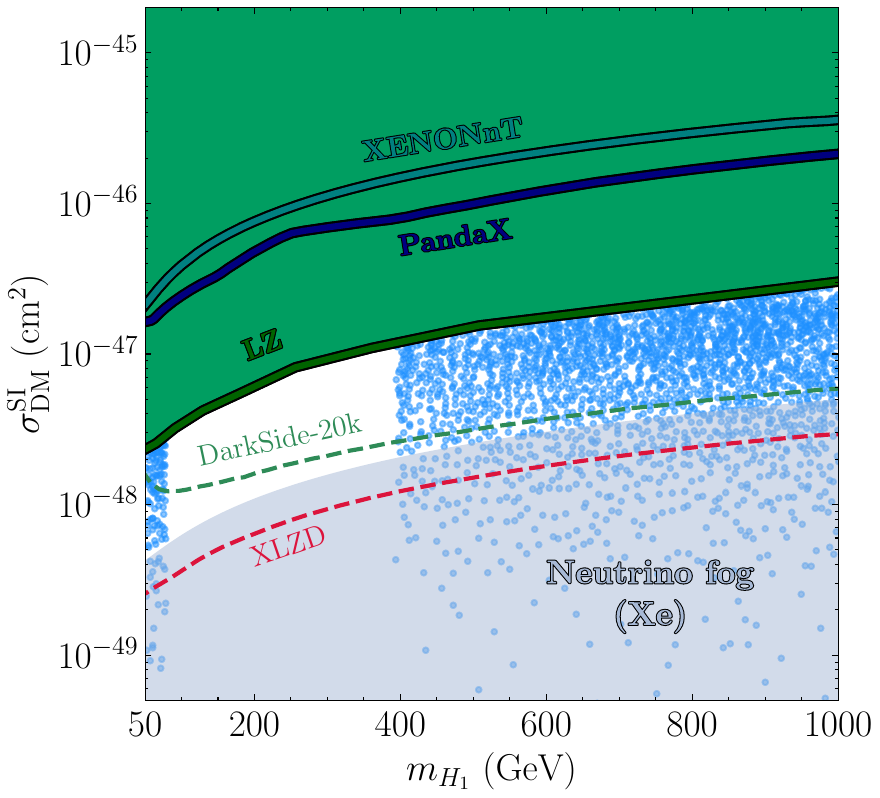}
	\captionsetup{font=small}
	\caption[Introductory plane of $\sigma_{H_1}^{SI}$ and $m_{\textrm{DM}}$ with colour code for $g_1$]{Plane of the direct detection cross-section $\sigma_{H_1}^{SI}$ and $m_{\textrm{DM}}$. The points in blue pass relic bounds exactly, all theoretical and experimental constraints. The lines shown are the experimental upper bounds from XENONnT~\cite{XENON:2025vwd}, PandaX-4T~\cite{PandaX:2024qfu} and LZ~\cite{LZ:2024zvo}. In dashed lines, the pojections for the next generation experiments DarkSide-20k~\cite{DarkSide-20k:2017zyg} and XLZD~\cite{XLZD:2024nsu} experiments. The neutrino fog~\cite{OHare:2021utq} is shown as the grey shaded region.}
	\label{fig:Z3-DD}
\end{figure}

Following Fig.~\ref{fig:Z3-NoCB-colorcode}-right, the benchmark scenario G allows for valid points in the region $420 \leq m_{\textrm{DM}} \leq 480\, \textrm{GeV}$, comparing with Figure 3 of Ref.~\cite{Aranda:2019vda}. We see the same trend for favoured values of the coupling $g_1$, which, again, are symmetric in our study. This range can be expanded to $m_{\textrm{DM}} \lesssim 700\, \textrm{GeV}$ when also considering Figure 6 of Ref.~\cite{Aranda:2019vda}.

After imposing all the required BFB conditions, we now scan the full mass range in the same limit $\theta = \frac{\pi}{4}$, with a vanishing  $ZH_iA_i \sim \cos(2\theta)$ vertex. In Fig.~\ref{fig:Z3-DD}, we show the results of scans together with the direct detection experimental limits from LZ in blue~\cite{LZ:2024zvo}, the prospects of XLZD in red~\cite{DARWIN:2016hyl,XLZD:2024nsu} and the neutrino fog in grey, along with the older experimental limits. These points are in full agreement with the relic density bounds and \texttt{HiggsTools}~\cite{Bahl:2022igd}. In these plots, we see how it is possible to populate the high mass range of the model, which was not covered in Ref.~\cite{Aranda:2019vda}. The coupling $g_1$ is extremely constrained by the current bounds, yet it is not excluded by future experiments with points inside the neutrino fog region, as described in Refs.~\cite{Billard:2013qya,OHare:2020lva,OHare:2021utq}. 

\begin{figure}[h!]
	\centering
	\includegraphics[width = 0.98\textwidth]{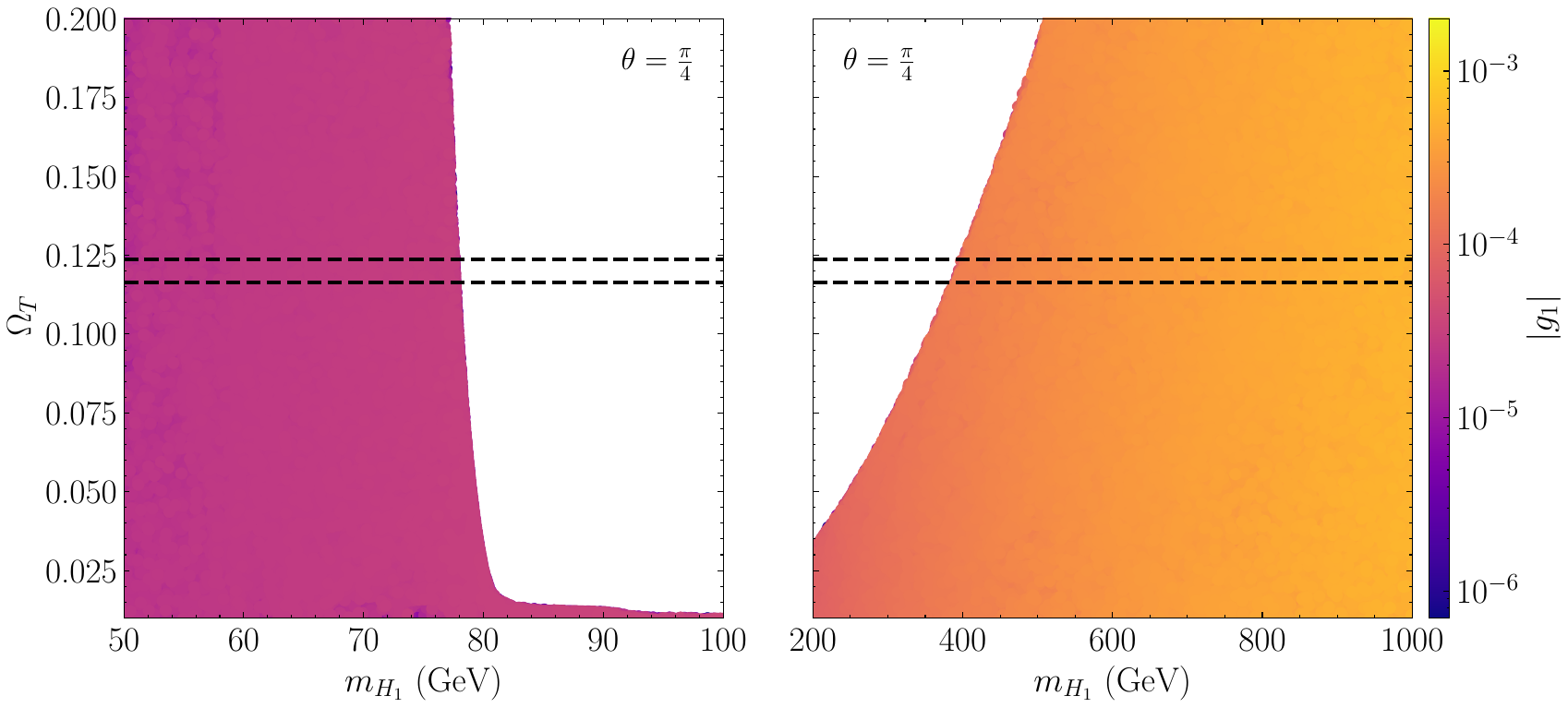}
	\captionsetup{font=small}
	\caption{Results obtained for the limit $\theta = \frac{\pi}{4}$ without fixing any of the variables in Eq.~\eqref{eq:massvariables}, with colour code for $|g_1|$. All necessary theoretical constraints are applied, along with the relevant updated bounds from \texttt{HiggsTools} and direct detection. The horizontal dashed lines are the bounds from relic density, considered at $3\sigma$. These points are sampled using \emph{seeds} and \emph{focus} on the parameters and observables of interest shown.}
	\label{fig:Z3-final}
\end{figure}

In order to improve the coverage of the parameter space, we follow the strategy of extracting \textit{seeds} from the points in Fig.~\ref{fig:Z3-DD} to be used for subsequent runs. Combining this with \emph{novelty reward} on the parameters and observables of interest, such as $m_{\textrm{DM}}$, $g_1$ and $\Omega_T$, we can explore these planes efficiently. Shown in Fig.~\ref{fig:Z3-final}, we observe that it is possible to cover the entire high mass region above $m_{\textrm{DM}} \gtrsim 386 \textrm{ GeV}$ while satisfying every constraint. 

We point out that the current indirect detection constraints pose only mild exclusions for this model, similar to the $\Z2 \times \Z2'$ 3HDM of Ref.~\cite{Boto:2024tzp}, and the dominant channel is also $H_1 H_1/ A_1 A_1 \rightarrow b \bar{b} $ for lower masses. With increasingly higher DM masses, the contribution from the $b\bar{b}$ process decreases, while the decay to $W^{+*} W^{-*}$ becomes dominant. Near $m_{\textrm{DM}} \sim m_W$, the process $H_1 H_1/ A_1 A_1 \rightarrow W^{+} W^{-}$ depletes the relic sharply, thus resulting in the observed steep boundary around this region. Direct detection constraints exclude larger values of $g_1$, thereby limiting the coupling in this low-mass regime to be of order $\mathcal{O}(10^{-5})$. There is also an increase in difficulty when sampling points for $m_{\textrm{DM}} \lesssim m_h / 2 \textrm{ GeV}$, as this region is extremely constrained due to bounds from the $125\,\textrm{GeV}$ Higgs boson invisible branching ratio, measured by the ATLAS experiment at $\textrm{BR}(h_{125} \rightarrow \textrm{invisible}) \leq 0.107 \pm 0.077$~\cite{ATLAS:2023tkt}.

For the intermediate mass region of $m_{\textrm{DM}} > m_W$ up until $m_{\textrm{DM}} \lesssim 386 \,\textrm{GeV}$, DM is under-produced due to the annihilation into gauge bosons. From this point upward, until our imposed scan upper mass $m_{\textrm{DM}} \leq 1000 \, \textrm{GeV}$, it is possible to account for the full relic in the entire mass range, while obeying every constraint, with a $g_1$ coupling of the order $\mathcal{O}(10^{-4})$, increasing with the DM mass. 

\subsection{Other solutions to \texorpdfstring{\boldmath$\cos(2\theta)=0$}{}}

We complete the analysis of the $\Z3$ 3HDM as a  DM model by considering the alternative solution to the condition of a vanishing vertex $ZH_iA_i \sim \cos(2\theta)$. As there are processes with a dependence on $\sin(2\theta)$, as shown in the diagram, the physical results do change when taking a value of $\theta$ of $-\tfrac{\pi}{4}$. 

\begin{figure}[H]
    \centering
    \begin{minipage}[c]{0.38\textwidth}
        \centering
        \begin{tikzpicture}
            \begin{feynman}
                \vertex (v) at (0,0);
                \vertex (h) at (-2,0) {\(h\)};
                \vertex (H1a) at (1.2,1.4) {\(H_1\)};
                \vertex (H1b) at (1.2,-1.4) {\(H_1\)};

                \diagram*{
                    (h) -- [scalar] (v),
                    (v) -- [scalar] (H1a),
                    (v) -- [scalar] (H1b),
                };
            \end{feynman}
        \end{tikzpicture}
    \end{minipage}%
    \hspace{-1cm}%
    \begin{minipage}[c]{0.58\textwidth}
        \begin{equation}
            \begin{aligned}
                g_1 ={}&
                (\lambda_5+\lambda_8)\cos^2\theta
                +(\lambda_6+\lambda_9)\sin^2\theta +\lambda_{12}\sin(2\theta)\,.
            \end{aligned}
            \label{eq:h-H1-H1-coupling}
        \end{equation}
    \end{minipage}
\end{figure}

\begin{figure}[h!]
	\centering
	\includegraphics[width = 0.65\textwidth]{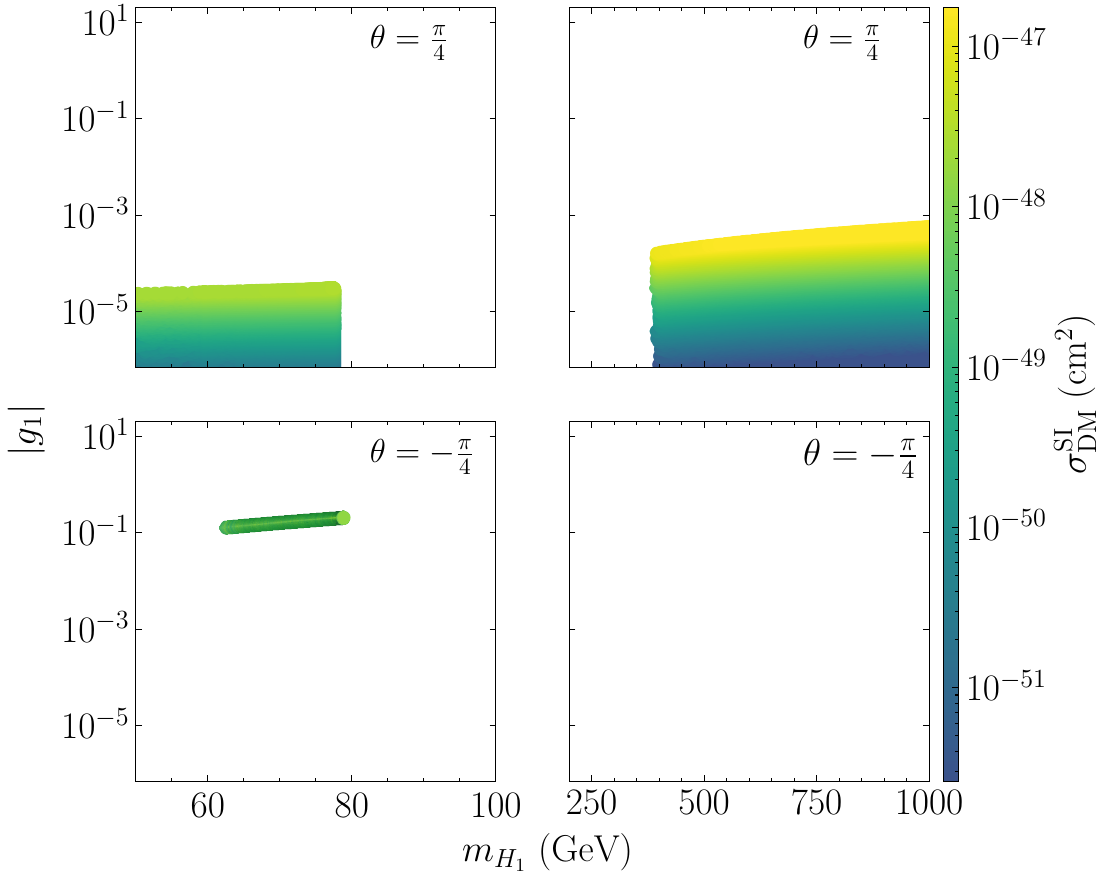}
	\captionsetup{font=small}
	\caption{Four plots showing the DM-higgs coupling, the DM mass and the direct detection cross-section. The top left and top right have $\theta = \tfrac{\pi}{4}$, and the bottom left and bottom right have $\theta = -\tfrac{\pi}{4}$. These points meet all the imposed constraints, including relic bounds. }
	\label{fig:theta_region_low_mass}
\end{figure}

We summarize the results obtained with the two different $\theta$ choices in Fig.~\ref{fig:theta_region_low_mass}, showing points that simultaneously meet all existing constraints and predict the correct relic density. In sampling the scenarios shown, no parameter of the model was fixed other than the angle $\theta$.  For the case with an heavy DM candidate, the case of $-\tfrac{\pi}{4}$  does not present viable solutions in the scans performed. For lower mass scenarios, there is a region of mass $62.5~\textrm{GeV} \lesssim m_{\textrm{DM}} \lesssim m_W$ with points meeting all imposed constraints. Additionally, we consistently observe that the DM-higgs coupling is of order $\mathcal{O}(0.1)$, significantly higher than in the previously discussed scenario of $\theta=\tfrac{\pi}{4}$, of at most $\mathcal{O}(10^{-4})$.

\section{Conclusions}\label{sec:conclusions}

We have conducted a detailed study of a viable particle dark matter model with two
mass-degenerate states of opposite CP quantum number, $H_1$ and $A_1$. The SM is extended by two additional Higgs doublets, and a discrete $\Z3$ symmetry acting on the three doublet fields ensures the stability of the dark matter states. Starting from the scalar potential, we reviewed the derivation of the potential parameters in terms of the independent parameters, including the physical masses in Eq.\eqref{eq:independentparameters}. We then proceeded to describe the theoretical consistency conditions that the model must satisfy, with the boundedness from below, global minimum and unitarity, as well as the relevant experimental constraints from existing searches.

The conditions obtained with the method described for boundedness from below are only sufficient, not necessary. A point may therefore fail them and still admit no runaway direction. In such a case, the sampling procedure would discard a physically valid point, potentially missing important physics. We introduce a method to numerically study the boundedness from below of points that fail the sufficient conditions but satisfy the (partial) necessary conditions along neutral directions. 

To explore the parameter space of the model, we employed machine learning techniques capable of efficiently probing complex, high-dimensional BSM scenarios without requiring pre-existing training data. Specifically, we combined an Evolutionary Strategy algorithm~\cite{deSouza:2022uhk} with an anomaly-detection-based Novelty Reward mechanism~\cite{Romao:2024gjx} to identify phenomenologically viable points. 

Our analysis was structured in three parts. First, we followed the approach of Ref.~\cite{Aranda:2019vda} by imposing the conservative limit $\theta = \frac{\pi}{4}$, which eliminates the $ZH_iA_i$ coupling (proportional to $\cos 2\theta$) capable of depleting the relic abundance. Within two benchmark scenarios defined by fixed differences between the physical masses, we successfully reproduced the previous results under the corresponding constraints. However, upon applying our more stringent theoretical consistency conditions, notably the boundedness from below, the scanned points were found to be excluded. This outcome highlights the critical importance of a robust analytical foundation in the study of BSM models.

We go beyond the literature by allowing all physical masses to vary freely within the scan range defined in Eq.~\eqref{eq:scanparameters}, while retaining the conservative choice $\theta = \frac{\pi}{4}$. In this condition, we find that the relic abundance requirement can be satisfied simultaneously with all theoretical and experimental constraints, yielding two distinct allowed regions for the mass of the two degenerate DM states: a low-mass region $50~\textrm{GeV} \lesssim m_{\textrm{DM}} \lesssim m_W$, and a high-mass region extending from $m_{\textrm{DM}} \gtrsim 386~\textrm{GeV}$ up to the scan limit of $1000~\textrm{GeV}$. This broadening of the viable spectrum on both ends is made possible by dedicated searches. In the intermediate mass region, DM is found to be under-produced due to the efficient annihilation into gauge bosons. The stringent direct detection constraints restrict the Higgs-DM coupling $g_1$ to values of order $10^{-5}$ in the low-mass region and $10^{-4}$ in the high-mass region. Notably, many of the viable points we identify still lie within the neutrino fog, where it presents an irreducible background. As a result, even next-generation direct detection experiments may not be able to definitively rule out the model.

Finally, we consider an alternative solutions to a vanishing vertex $ZH_iA_i \sim \cos(2\theta)$. For the case of $\theta=-\tfrac{\pi}{4}$, we find that agreement with all constraints can be achieved with significantly larger Higgs-DM couplings, reaching values as high as $|g_1| \sim 0.1$.
To address the computational demands of the scan, we develop density-based methods to automatically select prototype points that serve as seeds for subsequent scan runs, following the procedures introduced in Section~\ref{subsec:Prototype} and detailed in Appendix~\ref{app:appendixB}. The methodology described can be applied to any problem of constrained sampling. The diversity metrics introduced in Appendix~\ref{app:appendixB} provide a first quantitative comparison of these methods, showing them to be complementary rather than redundant. A full efficiency study in terms of coverage per CPU-hour and convergence rate to new regions is left for future work, as is the use of such metrics to fully automate the seed-selection step. Another promising application of prototype selection lies in identifying benchmark parameter choices that maximize a specific signature for a given model. For example, those most likely to be probed at upcoming collider experiments or in next-generation direct detection DM searches.

\section*{Acknowledgments}

We thank João Paulo Silva for discussions.
This work is supported in part by the Portuguese Fundação para a Ciência e Tecnologia (FCT) through the PRR (Recovery and Resilience Plan), within the scope of the investment ``RE-C06-i06 - Science Plus Capacity Building'', measure ``RE-C06-i06.m02 - Reinforcement of financing for International Partnerships in Science, Technology and Innovation of the PRR'', with a grant under the project with reference 2024.01362.CERN, and through Centro de Física Teórica de Partículas (CFTP) Contracts UIDB/00777/2020, and UIDP/00777/2020, partially funded through POCTI (FEDER), COMPETE, QREN, and the EU. The work of R.B. is supported by the Deutsche Forschungsgemeinschaft (DFG, German Research Foundation) under grant 396021762-TRR 257. MCR is supported by the STFC under Grant No. ST/T001011/1. F.A. de Souza is also supported by FCT under the research grant with reference
No. UI/BD/153105/2022.

\bibliographystyle{JHEP} 
\bibliography{ref}

\clearpage

\appendix

\section{Contributions to \texorpdfstring{\boldmath$\langle \sigma v \rangle$}{sigma V}}\label{app:appendixA}

We expected the same contribution from both DM components, $H_1$ and $A_1$, to the relic density. This can only be achieved with the appropriate method to calculate the relic, as explained in \ref{subsec:RelicDensity}, due to the processes of co-scattering of the form $\textrm{DM} \;\textrm{SM} \rightarrow \textrm{DM}' \; \textrm{SM}'$ included only in that method. 

\begin{figure}[htbp!]
	\centering
	\includegraphics[width = 0.4\textwidth]{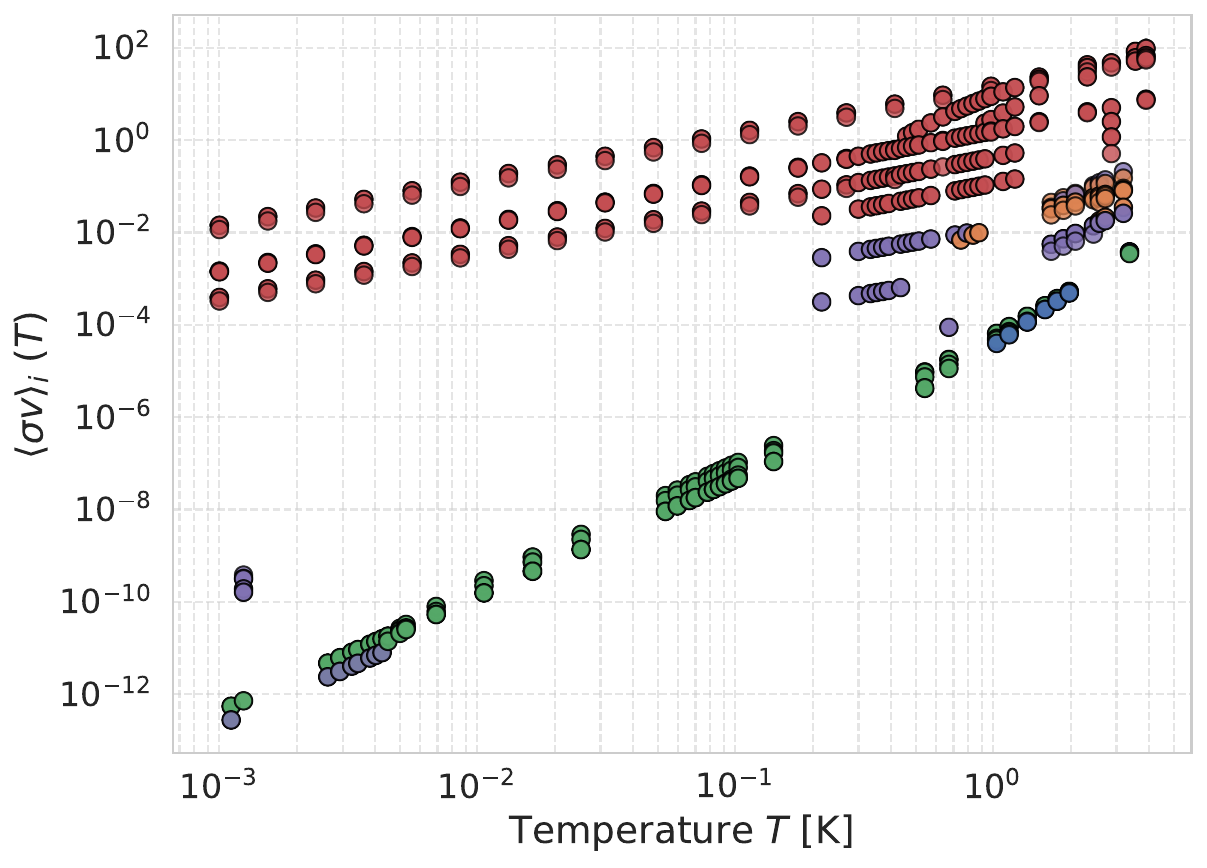}
    \hfill
	\raisebox{2.5cm}{\includegraphics[width = 0.15\textwidth]{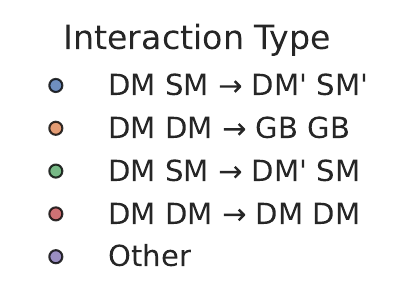}}
    \hfill
	\includegraphics[width = 0.4\textwidth]{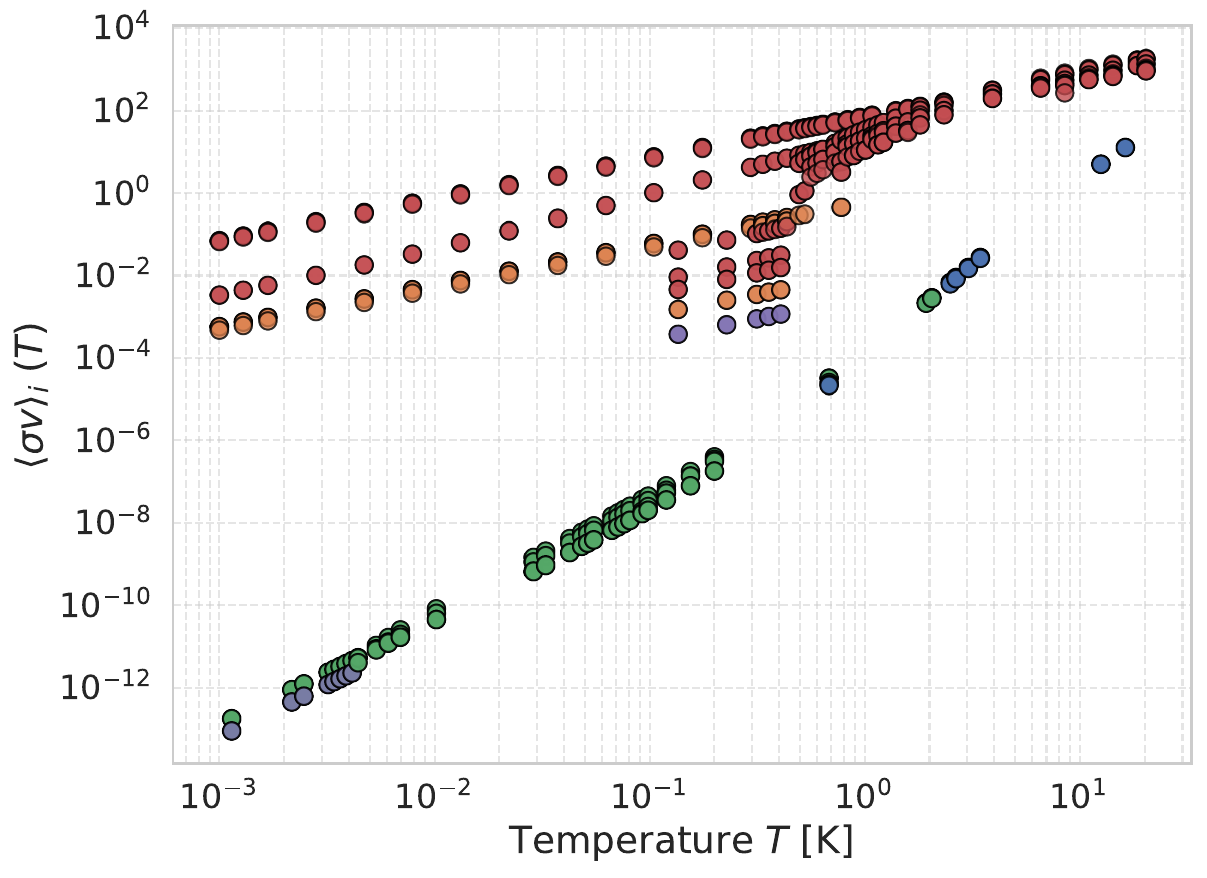}
	\captionsetup{font=small}
    \caption{Processes contributing to $\langle \sigma v \rangle$ for $m_{H_1} = 70\,\textrm{GeV}$ on the left and $m_{H_1} = 400\,\textrm{GeV}$ on the right, as a function of the temperature of the Universe. The legend distinguishes points based on the interaction type, either with DM species, gauge bosons (GB) and other SM particles. Co-scattering processes such as $H^{-}_2 \; u \rightarrow d \; H_1$, represented in the legend with the form $\textrm{DM}\; \textrm{SM} \rightarrow \textrm{DM'}\;\textrm{SM'}$, are relevant at high temperatures and contribute to deplete the relic.}
	\label{fig:contribution-sigmav-full}
\end{figure}

In Fig.~\ref{fig:contribution-sigmav-full}, we show the top contributions to the thermally-averaged cross-section times velocity, $\langle \sigma v \rangle$, with the expanding universe, presented as a function of the temperate, $T$. To follow the time evolution, $t$, of the universe, the plots should be read from right to left, since $T \propto t^{-1}$.

From both plots, we can conclude that the processes of co-scattering, displayed mainly in pink and purple lines, are more relevant for higher temperatures, and become increasingly less effective as the universe cools down. They are essential to preserve balance between the DM candidates and to obtain the correct relic, as it tends to drop significantly when including more processes. If the contributions of such processes are not included, one might expect to get the correct relic when in fact, it is under-abundant, as $\langle \sigma v \rangle \propto \Omega^{-1}$ for freeze-out.

\section{Prototype Selection Methods}
\label{app:appendixB}

Here we present the implementation details of the seed selection methods listed in Section~\ref{subsec:Prototype}. Specifically, we develop methods to automatically identify regions of the parameter space that warrant further exploration by employing multiple, distinct techniques, each characterised by its own strengths and limitations, thereby providing complementary approaches.

\subsection{Seed selection with HGBC in sparse regions}

The process of extracting seeds can be performed automatically with a machine learning classifier. For this work, we employed a Gradient Boosting (GB) algorithm~\cite{Freund:1997xna, Breiman:1999, Friedman:2021}, that uses decision trees as predictors. A GB builds an ensemble of decision trees which are sequentially fit on the residual errors made of the previous one. In this work, we employ a variation of GB known as Histogram-Based Gradient Boosting~\cite{10.1007/978-3-030-37334-4_4}. Namely, we use the Histogram Gradient Boosting Classifier (HGBC), that works by binning input features and replacing them with integers. With this implementation, it is possible to use more memory-efficient data-structures and it removes the need for sorting the features when training each tree.  As a trade-off, binning causes some precision loss that ends up acting as a regularizer, either reducing overfitting or causing some underfitting. In our scenario, this proves useful, as we want to select a representative sample from the various regions that may need seeds, and not necessarily collect a large sample from one specific region.

Here, we provide examples of how HGBC can be used for seed selection. It is trained to distinguish the data from overlapping random noise. Given that HGBC has achieved an acceptable performance in its classification task, we can then select some points to be used as seeds, typically $1\%$ or less from the most outlying quantile in the data. In practice, a sizeable amount of training data is required in order to achieve an acceptable classification performance. Thus, this method is best used after a sufficiently large number of valid points have already been obtained by CMA-ES.

\begin{figure}[htbp!]
	\centering
	\includegraphics[width = 0.32\textwidth]{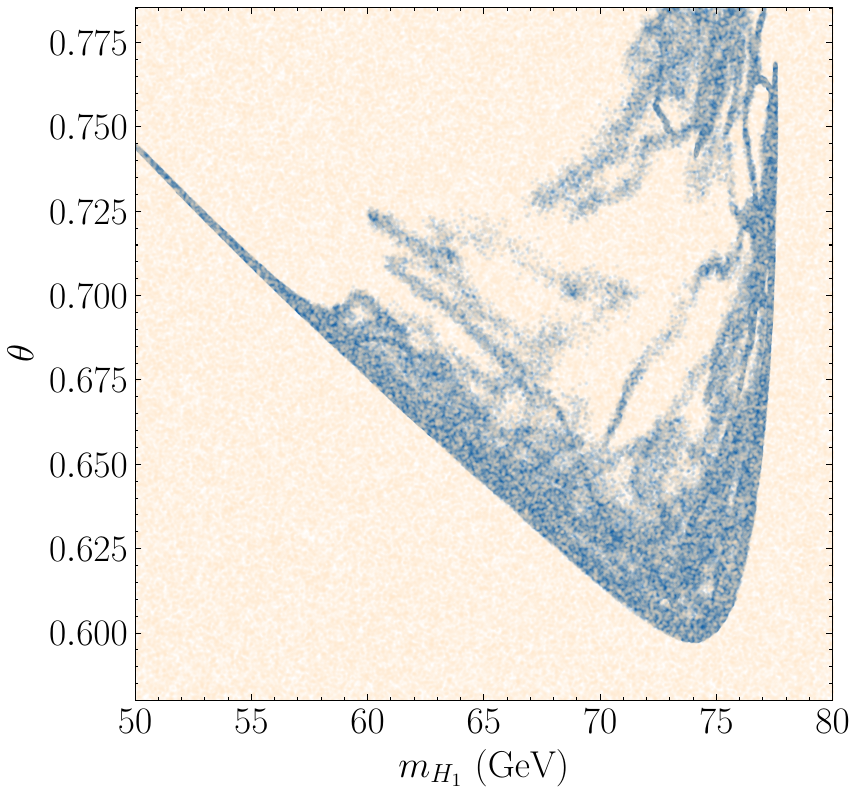}
    \hspace{2.5cm}
	\includegraphics[width = 0.32\textwidth]{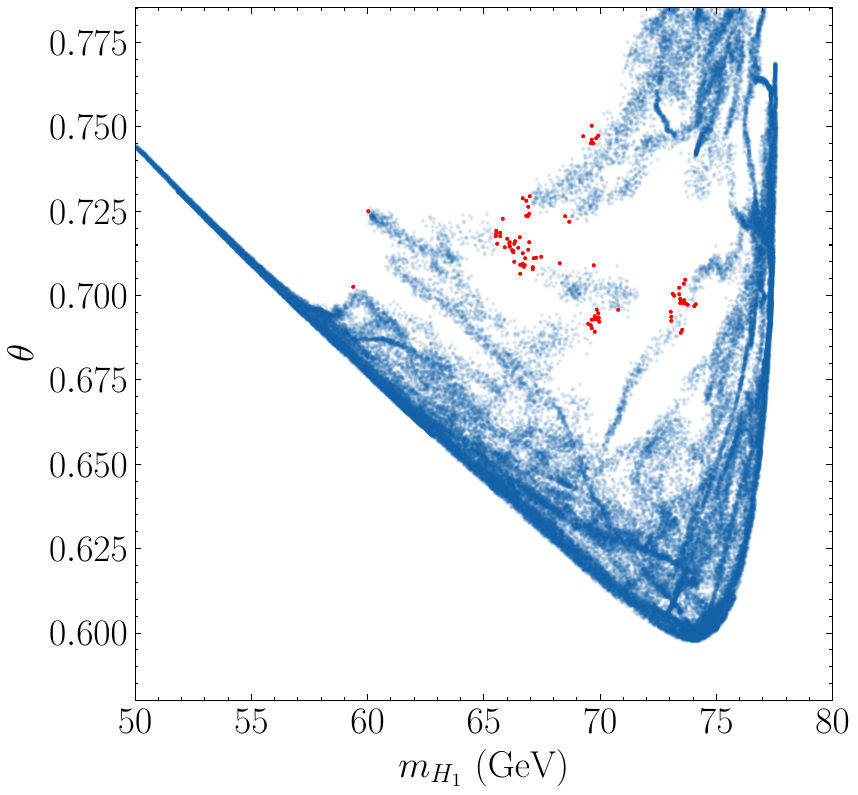}
	\captionsetup{font=small}
    \caption{On the left, the uniform noise mask overlapped with data to train HGBC. On the right, points selected from the least dense regions in red.}
	\label{fig:theta-vs-mass-seeds}
\end{figure}

To exemplify the use of the HGBC, in Fig.~\ref{fig:theta-vs-mass-seeds} we show a plane originating from initial runs, covered by noise data points, and the result of the selection on the right. In this case, we are keeping only the $0.001\%$ of points that are harder to distinguish from the uniform background. As we can see, this technique is especially capable of identifying regions of global lower density, where the user might want to focus further exploration.

\subsection{Data Clustering using Mini-Batch \texorpdfstring{\boldmath$k$}{KMB}-Means}

An early challenge encountered in this work was the generation of very large datasets of valid points. While this outcome attests to the effectiveness of the exploration methodology employed herein, it introduces substantial difficulties for subsequent data analysis and for the selection of new points to serve as seeds for additional runs. To address this issue, we employed a clustering algorithm to identify representative prototypes from the collected dataset of valid points, which were then used in downstream analysis and for seed selection. To this effect, we considered Mini-Batch $k$-Means (MBKM), a more efficient variation of the clustering algorithm $k$-Means which divides the full dataset into separate groups of equal variance, called clusters, while minimising a quantity know as the inertia. The inertia is simply the mean squared distance between each instance and its corresponding centroid, the cluster mean. MBKM uses batches instead of the full dataset, slightly adjusting the cluster centroids at each iteration while being much faster than regular $k$-Means.  For our purposes, clustering the collection of amassed valid points comes with two applications.

\begin{figure}[htbp!]
	\centering
	\includegraphics[width = 0.32\textwidth]{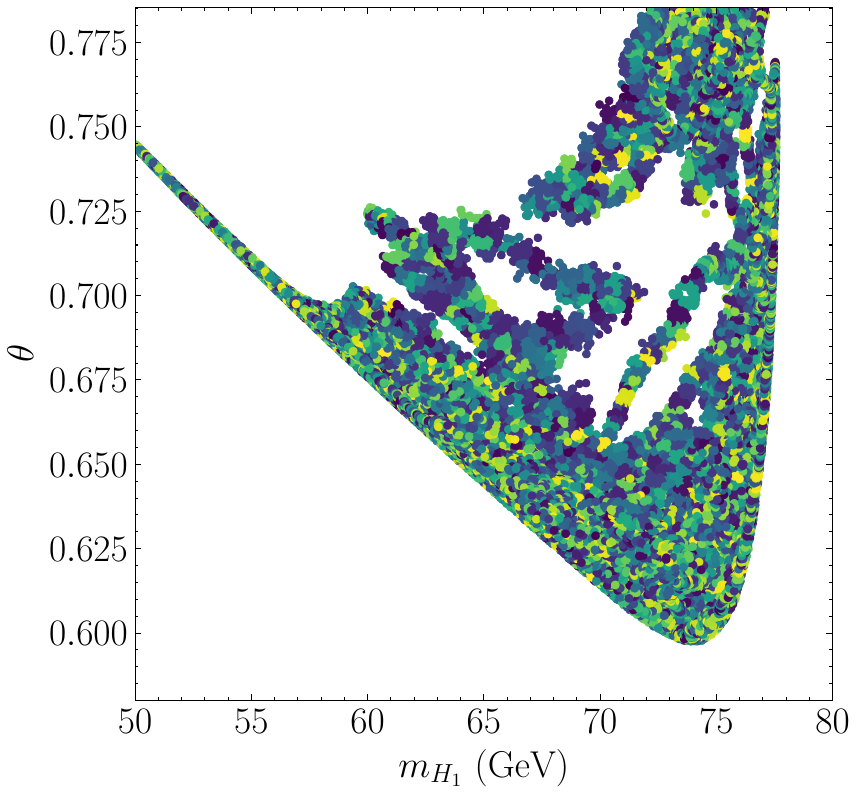}
    \hspace{2.5cm}
	\includegraphics[width = 0.32\textwidth]{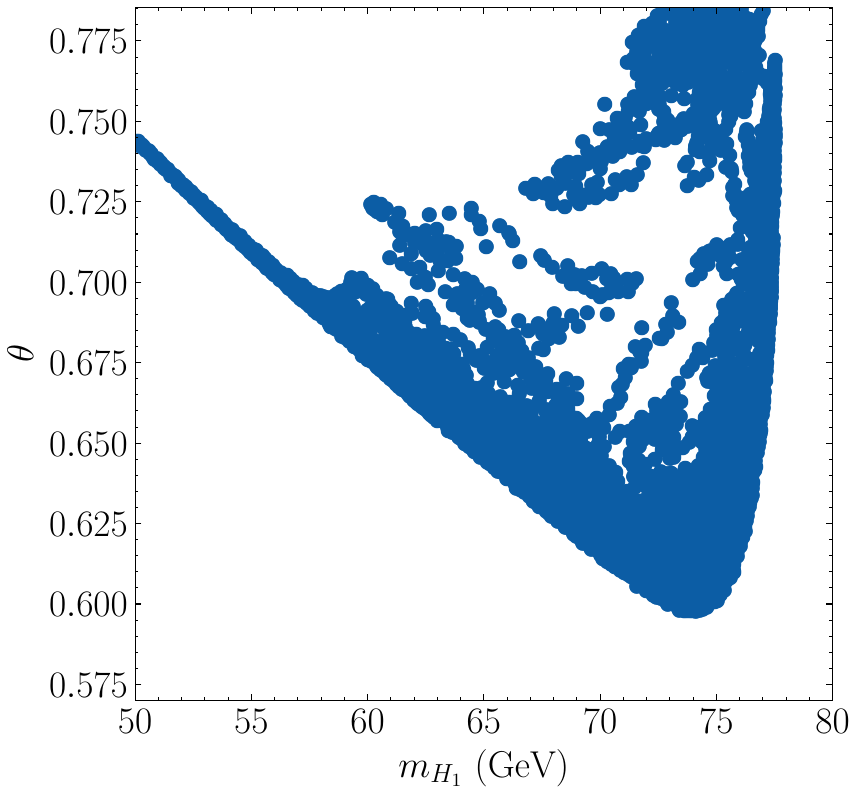}
	\captionsetup{font=small}
    \caption{On the left, the clusters defined by MBKM. In this case, as MBKM is being used for sampling, we define a large number of clusters. On the right, the results of the sampling by selecting points near the cluster centroids.}
	\label{fig:theta-vs-mass-kmb-sampling}
\end{figure}

A first application for MBKM is data condensation. Once the collection of valid points has grown considerably large ($\gtrsim 10^{6}$ points) the data analysis can start to become unpractical. In that case, it becomes useful to employ MBKM to provide a representation of the data around each cluster, reducing its size while maintaining the density distribution. On Fig.~\ref{fig:theta-vs-mass-kmb-sampling}, we show how MBKM can be used to extract a representative sample of points from the dataset. We train MBKM with a large number of clusters. Notice how denser regions tend to form more compact clusters, with smaller radii. Then, one can simply collect the nearest points to the cluster centroids and use them to represent the dataset efficiently for further analysis.

\begin{figure}[htbp!]
	\centering
	\includegraphics[width = 0.32\textwidth]{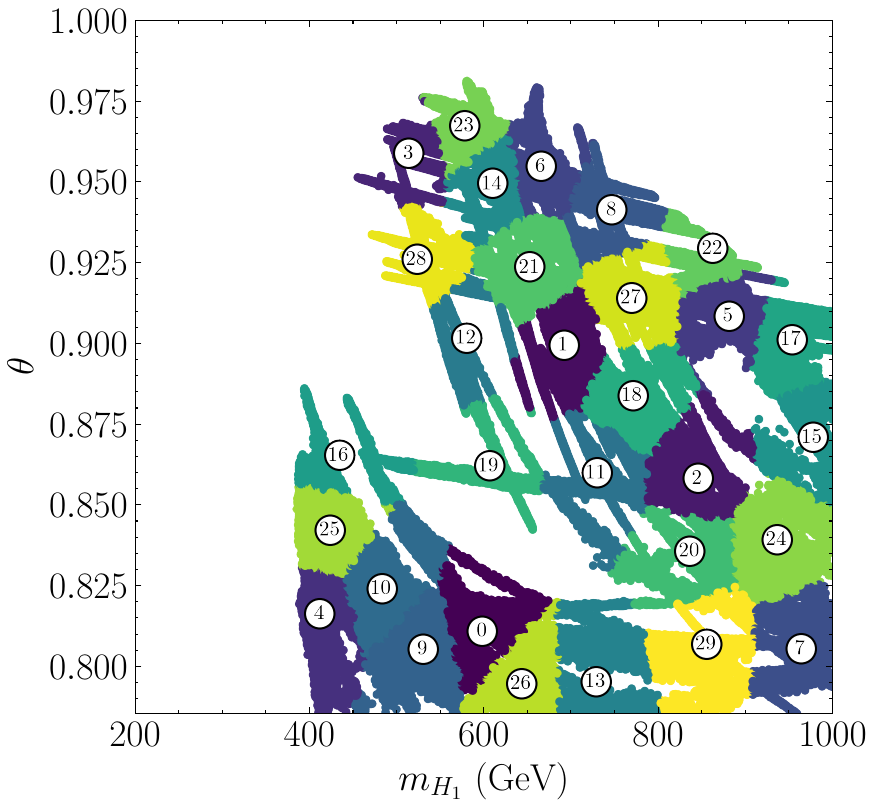}
    \hspace{2.5cm}
	\includegraphics[width = 0.32\textwidth]{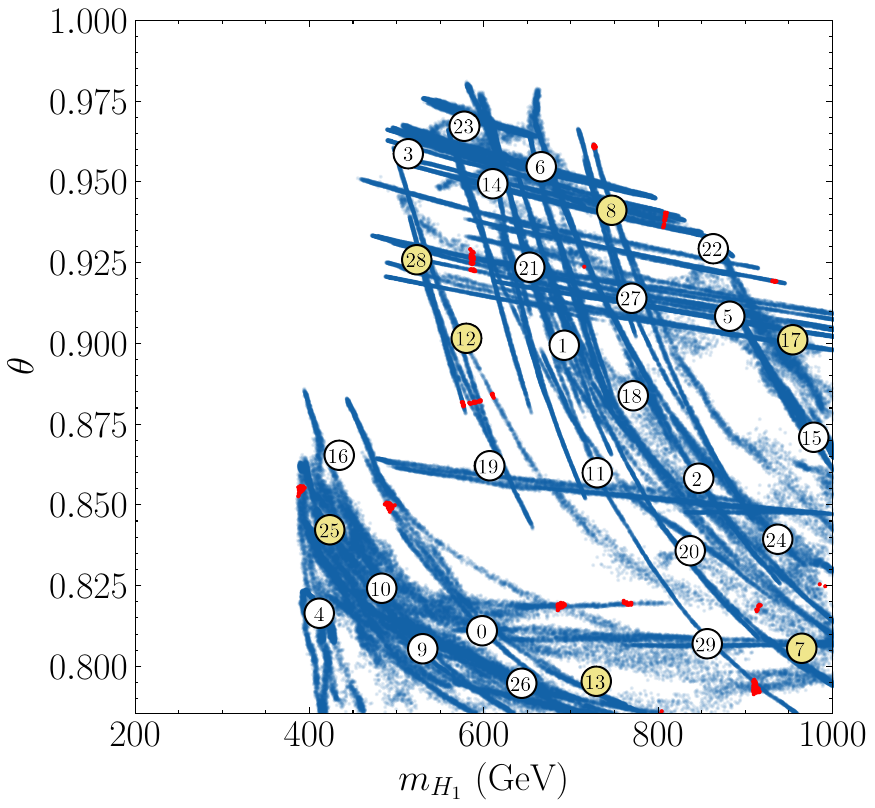}
	\captionsetup{font=small}
    \caption{On the left, the results from applying MBKM clustering for seed selection. On the right, the points in red are extracted from selecting the clusters $7, 8, 12, 13, 17, 25$ and $28$, highlighted in yellow, and isolating the ones that lay furthest from their respective centroids.}
	\label{fig:theta-vs-mass-kmb-seeds}
\end{figure}

A second application for MBKM is for seed selection, as points which lie furthest away from their corresponding centroids can be good candidates for seeds, since they are the least representative, or most outlier points from the clusters. This way, future CMA-ES scans using seeds from MBKM can work as effectively expanding the boundaries of the corresponding clusters. On Fig.~\ref{fig:theta-vs-mass-kmb-seeds}, we show how we can select seeds using MBKM. This can be done either by hand-picking the clusters we are interested in or done automatically, by selecting the $N$ clusters with higher radius. These can be combined to yield the desired results. Moreover, to restrict the selection to certain areas of the parameter space, it is possible to apply a user-defined rule that forces selection to a specific region of interest. It is important to note that in order to get the desired effect from MBKM, one needs to properly scale the data, for example with the Standard Scaler method, as to avoid directional biases during clustering.

Unlike HGBC, MBKM is a unsupervised learning technique, and so it is not limited to the amount of valid points amassed. However, MBKM assumes similar size, similar density and spherical shapes among clusters, which is not always the case, as can be seen for example with clusters 16 and 19 on Fig.~\ref{fig:theta-vs-mass-kmb-seeds}. Additionally, the number of clusters needs to be specified as an input parameter, which can make automatisation difficult for seed selection.

\subsection{Seed selection with Local Outlier Factor}

A third approach to automatically identify promising seeds used in this work was to use Local Outlier Factor (LOF). LOF is an algorithm specifically designed for outlier detection. Thus, its implementation is more straightforward than the two previous methods for seed selection. It works by identifying instances with local density that deviate significantly from their $k$-nearest neighbours. This way, it excels in isolating local outliers which are suitable to be used as seeds for CMA-ES. The $\sim 1\%$ most outlying valid points are good candidates for seeds. Unlike HGBC, LOF is a unsupervised learning technique, does not require a large training data and does not require a uniform background. Unlike MBKM, LOF does not assume a specific shape for its local neighbourhoods. The number of neighbours needs to be specified by the user, however $k=20$ neighbours usually works well.

\subsection{Complementary seed selection methods}

\begin{figure}[htbp!]
	\centering
	\includegraphics[width = 0.32\textwidth]{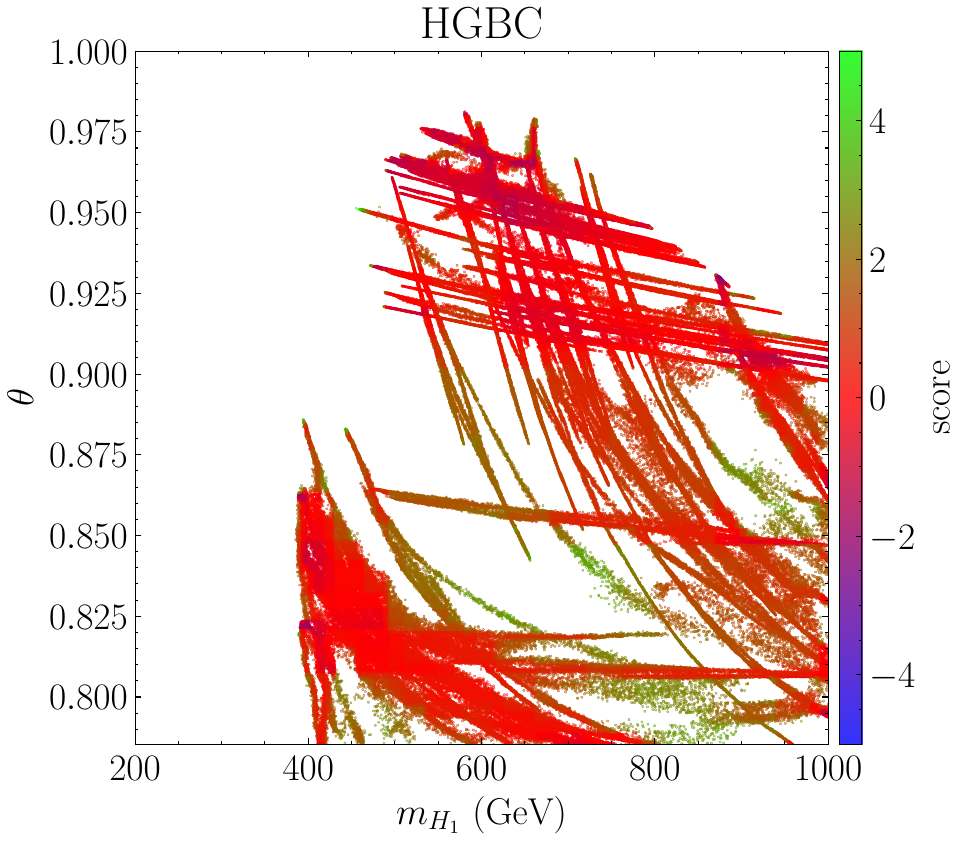}
	\hfill
	\includegraphics[width = 0.32\textwidth]{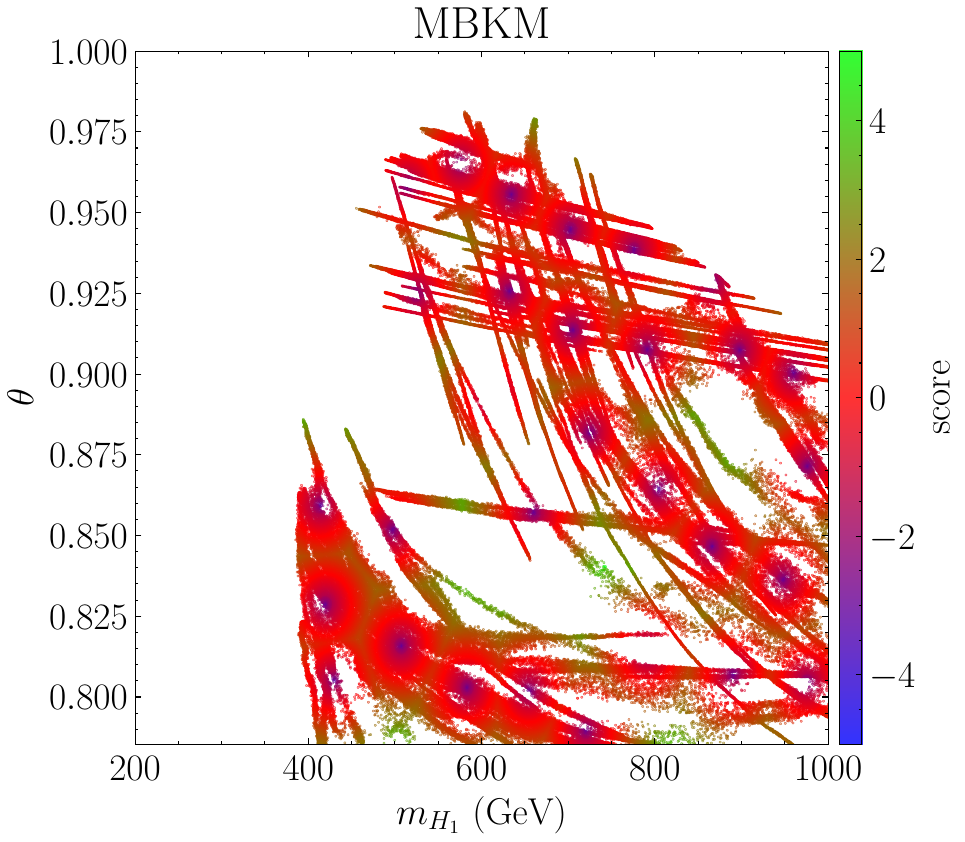}
	\hfill
	\includegraphics[width = 0.32\textwidth]{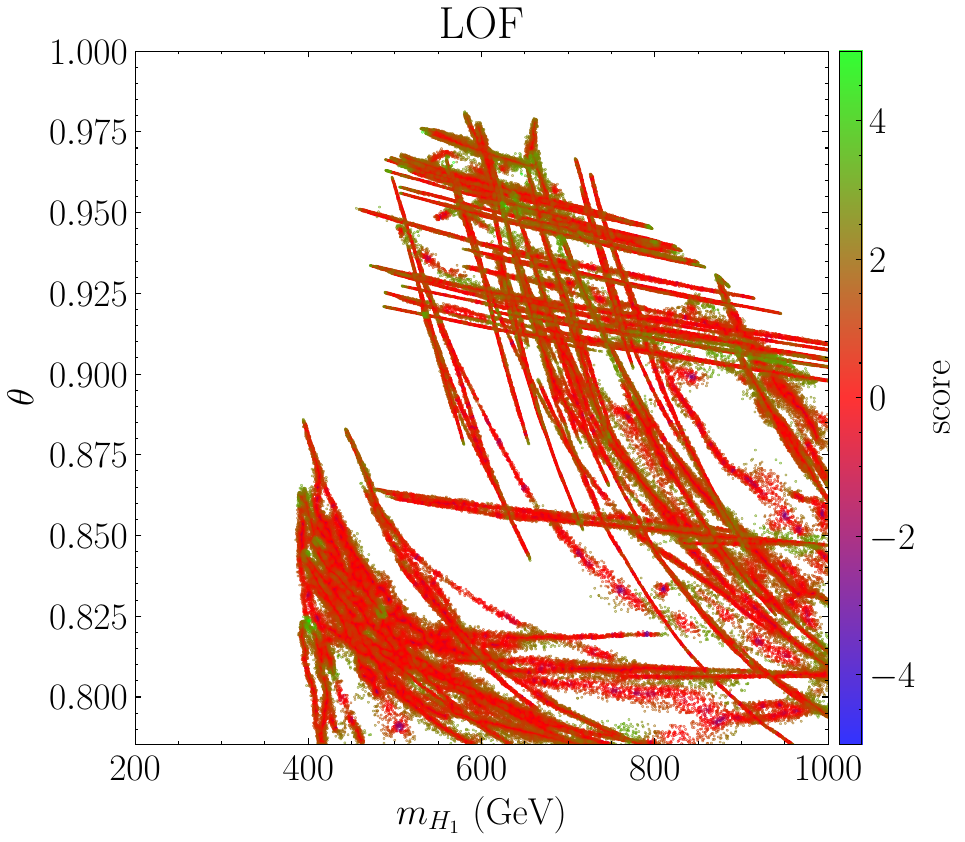}
	\captionsetup{font=small}
    \caption{Comparison between three different seed selection methods: HGBC (left), MBKM (center), and LOF (right). The points are colour coded by the score given by each method. For ease of comparison, the scores are normalised by a quantile transformation with a Gaussian distribution. The seeds are to be selected amongst the highest scoring (green) points.}
	\label{fig:theta-vs-mass-seeds-scores-comparison}
\end{figure}

\begin{figure}[htbp!]
	\centering
	\includegraphics[width = 0.32\textwidth]{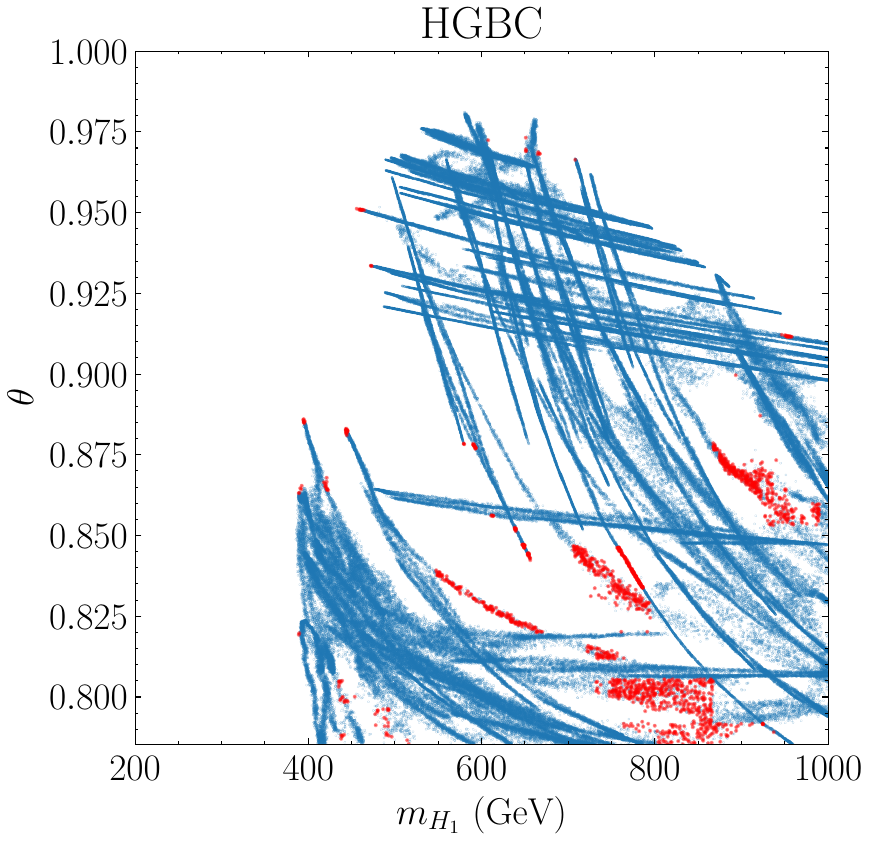}
	\hfill
	\includegraphics[width = 0.32\textwidth]{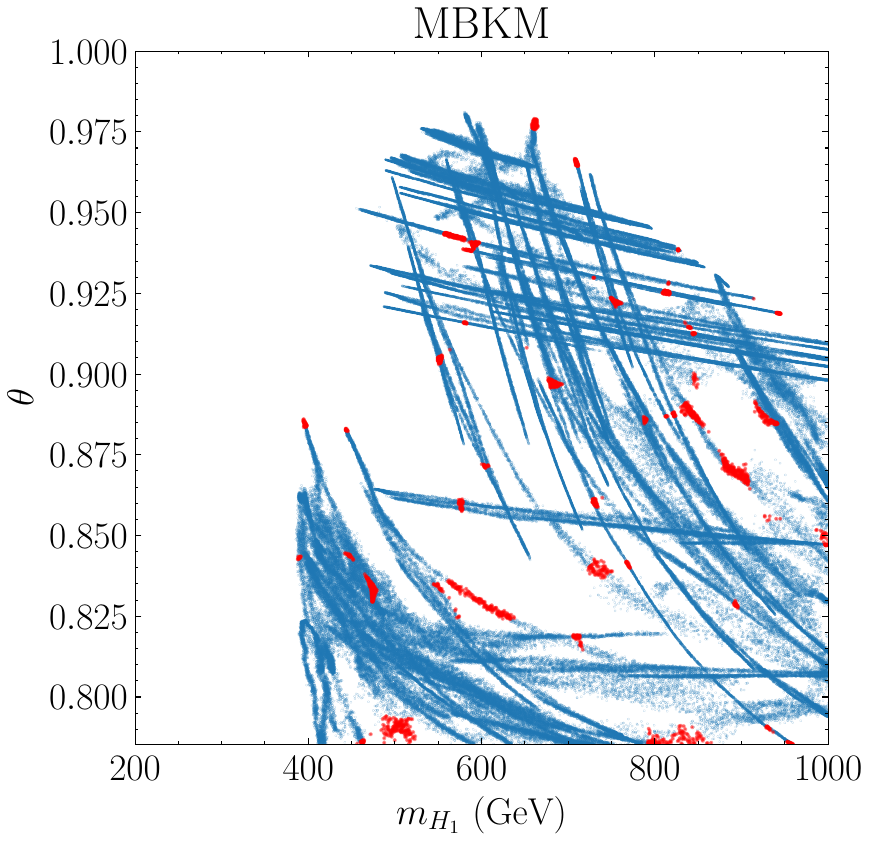}
	\hfill
	\includegraphics[width = 0.32\textwidth]{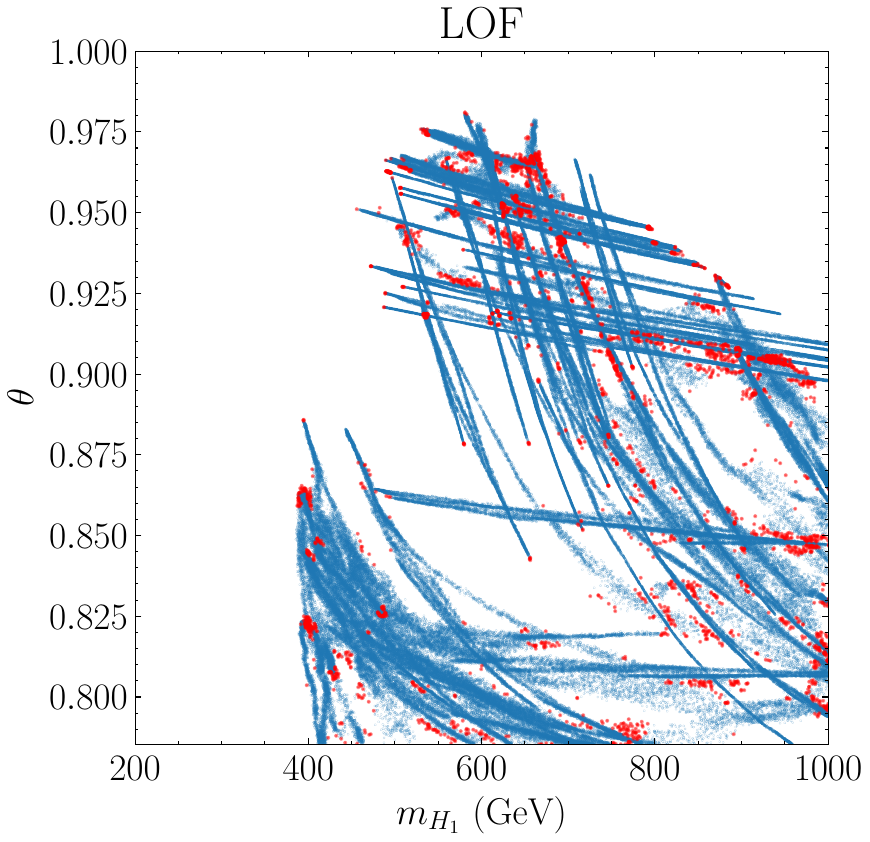}
	\captionsetup{font=small}
    \caption{Comparison between three different seed selection methods: HGBC (left), MBKM (center), and LOF (right). The red points represent the 1\% quantile with the highest scored points for each method.}
	\label{fig:theta-vs-mass-seeds-quantile-comparison}
\end{figure}

In order to demonstrate how the three methods -- HGBC, MBKM, and LOF -- differ, we performed a comparative analysis among them. An illustrative 
example is provided in Fig.~\ref{fig:theta-vs-mass-seeds-scores-comparison}. For a fair comparison, each method is trained/fitted in the same set of points. These points were obtained from the scans with $\theta \geq \frac{\pi}{4}$ and novelty reward focus on $m_{H_1}$ and $\theta$. The scores are normalised using a quantile transformation with a Gaussian distribution where the seeds are to be selected from amongst the points with the higher score values. The points which represent the 1\% highest scoring quantile for each seed selection method are shown in Fig.~\ref{fig:theta-vs-mass-seeds-quantile-comparison}.

For HGBC a higher score is associated with points which are harder to distinguish from random noise. We can see in Fig.~\ref{fig:theta-vs-mass-seeds-scores-comparison} the red-coloured horizontal and vertical slices corresponding to the tree partitions of the classifier containing the points which the classifier can easily isolate from the random noise. Whereas the green points with a higher score are harder to isolate from random noise with these tree partitions. This way, HGBC tends to prefer seeds in uniformly distributed and isolated areas, representing globally sparse
regions, as shown in the left plot of Fig.~\ref{fig:theta-vs-mass-seeds-quantile-comparison}.

For MBKM high scored points are the ones that lie furthest away from their cluster centroid. The colour pattern in Fig.~\ref{fig:theta-vs-mass-seeds-scores-comparison} clearly show this behaviour, where the blue points are the ones closer to the centroids, while the green ones are the farthest and the ones which the seeds ought to be selected from. Being so, this method can often select seeds that lie in the frontier of clustered regions and the intersection between them. This can be seen in the central plot in Fig.~\ref{fig:theta-vs-mass-seeds-quantile-comparison}.

LOF gives a higher score for points with a lower local density in a given neighbourhood. The colour pattern in Fig.~\ref{fig:theta-vs-mass-seeds-scores-comparison} confirms this strategy where the red points corresponds to inlier points in the local neighbourhood and the green points the outliers ones. Thus, LOF typically selects seeds amongst points lying in the frontier of these neighbourhoods and the intersection between them, corresponding to locally underdese regions, as illustrated in the right plot in Fig.~\ref{fig:theta-vs-mass-seeds-quantile-comparison}.

\subsection{Prototype-Selection Diversity Metrics}

It is possible to define diversity metrics that give a quantitative view of the different methods. For this study, we define the \textit{Coverage}, \textit{Separation} and \textit{Frontier} criteria. Each criterion captures a different notion of diversity and is normalised by comparison with random seed selections of the same size, so a ratio of $1$ corresponds to random performance.

Operationally, these methods proved essential throughout this work: they allowed successive scans to be compounded from existing valid points and targeted at specific regions on demand, whereas random seeding largely reconverged to already-populated regions and manual selection became impractical for datasets of $\gtrsim 10^{6}$ valid points.

Let
\begin{equation}
X = \{x_i\}_{i=1}^{N}
\end{equation}
be the full candidate set in scaled \((m_{H_1}, \theta)\) space with z-score standardization and let
\begin{equation}
S_m \subset X
\end{equation}
be the seed set selected by method \(m\), and let \(B_j\), \(j = 1, \dots, J = 25\), denote random baseline sets, uniformly sampled without replacement, each containing the same number of seeds as \(S_m\).

\textit{Coverage} measures how well the selected seeds represent the full candidate region. For each candidate point, the distance to the nearest selected seed is computed. The 95th percentile of these distances is used as a fill-distance measure. Smaller values indicate better coverage. The \textit{Coverage} ratio is defined as the random baseline fill distance divided by the method fill distance, so that values above $1$ indicate better global coverage than random and values below $1$ indicate poorer coverage. Since seed-selection methods are deliberately more selective than random, we expect $R_{\mathrm{cov}} < 1$; the desirable behaviour is to remain as close to $1$ as possible from below, i.e. to degrade global coverage as little as possible while being selective.

\textbf{Coverage:}

We start from the distance of each candidate point to its nearest selected seed,
\begin{equation}
d(x, S_m) = \min_{s \in S_m} \lVert x - s \rVert_2
\end{equation}
from which the fill distance of the full candidate set follows as its $95$th percentile,
\begin{equation}
Q_{95}(S_m) = \operatorname{percentile}_{95}
\left(
\{d(x, S_m): x \in X\}
\right)
\end{equation}
leading to the coverage ratio
\begin{equation}
R_{\mathrm{cov}}(S_m) =
\frac{
\frac{1}{J}\sum_{j=1}^{J} Q_{95}(B_j) 
}{
Q_{95}(S_m)
} \ .
\end{equation}

Lower fill distance is better, so the ratio is inverted. If \(R_{\mathrm{cov}}>1\), the method covers the full candidate region better than random.

\textit{Separation} measures redundancy among the selected seeds. For each selected seed, the distance to its nearest other selected seed is computed. The $5$th percentile of these nearest-neighbour distances is used as a lower-tail separation measure. Larger values indicate less local clumping. The \textbf{Separation} ratio is the method separation divided by the random baseline separation: values above $1$ indicate seeds that are more separated than random, while values below $1$ indicate more local clustering than random. As before, some clumping is unavoidable for a quantile-based selection, so the desirable behaviour is a ratio close to $1$ from below, i.e. seeds that cluster as little as possible given the selectivity of the method.

\textbf{Separation:}

For each selected seed, we first compute the distance to its nearest companion within the seed set,
\begin{equation}
\delta(s, S_m) =
\min_{\substack{t \in S_m \\ t \neq s}}
\lVert s - t \rVert_2
\end{equation}
and summarise the worst-case clumping of the seed set through the $5$th percentile of these distances,
\begin{equation}
P_{5}(S_m) =
\operatorname{percentile}_{5}
\left(
\{\delta(s, S_m): s \in S_m\}
\right)
\end{equation}
which enters the separation ratio as
\begin{equation}
R_{\mathrm{sep}}(S_m) =
\frac{
P_5(S_m)
}{
\frac{1}{J}\sum_{j=1}^{J} P_5(B_j)
}
\end{equation}
Since a larger $P_5$ means fewer near-duplicate seeds, the ratio is not inverted here: if \(R_{\mathrm{sep}} > 1\), the selected seeds are less locally redundant than a random selection of the same size, whereas \(R_{\mathrm{sep}} < 1\) signals local clumping.

\textit{Frontier} preference measures whether the selected seeds lie in sparse or boundary-like regions. For every candidate point, the distance to its \(K\)-th nearest neighbour in the full candidate set is computed. These distances are converted to percentile ranks, with higher percentiles corresponding to locally sparser regions. The frontier score of a seed set is the mean percentile rank of its selected seeds. The \textit{Frontier} ratio is the method frontier score divided by the random baseline frontier score, so values above $1$ indicate stronger preference for frontier or lower density regions than random. Contrary to the previous criteria, here the desirable behaviour is a ratio above $1$: seed-selection methods should actively target sparse, frontier-like regions more aggressively than random.

\textbf{Frontier:}

We first assign each candidate point a sparsity descriptor, given by the distance to its \(K\)-th nearest neighbour (here $K=20$) within the full candidate set,
\begin{equation}
\rho_i =
\text{distance from } x_i \text{ to its } K\text{-th nearest neighbor in } X.
\end{equation}
These distances are then converted to a percentile rank between 0 and 1 over the full candidate set,
\begin{equation}
F_i = \operatorname{percentile\_rank}(\rho_i),
\end{equation}
so that points in locally sparse regions receive $F_i$ close to $1$. The frontier score of a seed set is the mean rank of its selected seeds,
\begin{equation}
F(S_m) =
\frac{1}{|S_m|}
\sum_{x_i \in S_m} F_i
\end{equation}
which, when divided by the corresponding random baseline, yields the frontier ratio
\begin{equation}
R_{\mathrm{frontier}}(S_m) =
\frac{
F(S_m)
}{
\frac{1}{J}\sum_{j=1}^{J} F(B_j)
}.
\end{equation}
If \(R_{\mathrm{frontier}}>1\), the method selects more sparse/frontier-like points than random.

\begin{table}[htbp]
    \centering
    \begin{tabular}{lccc}
        \toprule
        & \textbf{Coverage} & \textbf{Separation} & \textbf{Frontier} \\
        \midrule
        \textbf{LOF} & $\mathbf{0.34}$ & $0.13$ & $1.66$ \\
        \textbf{HGBC} & $0.10$ & $\mathbf{0.29}$ & $\mathbf{1.89}$ \\
        \textbf{MBKM} & $0.08$ & $0.19$ & $1.68$ \\
        \bottomrule
    \end{tabular}
    \caption{Diversity metrics for the seed selection methods of this study, normalised to random selections of the same size (a ratio of $1$ corresponds to random performance). \textit{Coverage} quantifies a global representation of the candidate set, \textit{Separation} the local seed diversity and \textit{Frontier} the ability of the method to select seeds that lie on the outer edge of clusters. The best value in each column is shown in bold: closest to $1$ from below for \textit{Coverage} and \textit{Separation}, largest for \textit{Frontier}.}
    \label{tab:diversity-scores}
\end{table}

The final scores for each criterion on this specific training set are displayed in Tab.~\ref{tab:diversity-scores}. Such results indicate that LOF achieves the best \textit{Coverage}, HGBC the best \textit{Separation} and \textit{Frontier}, and MBKM, with an automatic selection, stands as an intermediate method. Its most advantageous use comes into play when selecting clusters by inspection, to explore specific regions of interest. We note that the three criteria are not independent: a strong preference for sparse regions, as quantified by \textit{Frontier}, necessarily degrades the fill distance and hence the \textit{Coverage} ratio, so the table effectively characterises a single trade-off between targeting sparseness and maintaining global coverage, together with the internal clumping of the seed set.

Comparing the seeds selected from each method in Fig.~\ref{fig:theta-vs-mass-seeds-quantile-comparison}, it is clear that HGBC, while excelling in isolating points in globally sparse regions, indicated in red on the left plot, struggles to identify points in highly populated areas which are surrounded by completely empty regions. We can see this explicitly for the points that lie in the stripe patterns with $\theta \gtrsim 0.9$, which HGBC fails to target. However, as indicated by the middle and right plots in Fig.~\ref{fig:theta-vs-mass-seeds-quantile-comparison}, LOF is able to target points in this region due to the focus on locality of its $k$-neighbours algorithm, which is absent in the HGBC case. This is also consistent with the metrics: by packing many seeds into these stripes, LOF attains the best \textit{Coverage} at the price of the lowest \textit{Separation}. MBKM enables the selection of specific regions of interest, which, when combined with parameter space scan rules to drive exploration, allows for a very efficient and targeted probing of the desired areas. If the goal is to efficiently and broadly populate the empty regions between the aforementioned stripes, LOF appears to be the most suitable choice. Finally, we can say the methods are complementary and seem to cover different use cases (global and local outliers) for targeting different interesting regions.

\end{document}